\documentclass[aip,pof,reprint]{revtex4-1}
\usepackage[dvipsnames,rgb,dvips]{xcolor}
\usepackage{graphicx}
\usepackage{float}
\usepackage{overpic}
\usepackage{amsmath,amssymb}
\usepackage{bm}
\usepackage{units}

\usepackage{array}
\newcolumntype{C}[1]{>{\centering\let\newline\\\arraybackslash\hspace{0pt}}p{#1}}

\newcommand{\lc}{\ensuremath{\varepsilon}}

\newcommand{\rd}{\ensuremath{\mathrm{d}}}

\newcommand\ve[1]{\boldsymbol{#1}}
\newcommand\ma[1]{\boldsymbol{#1}}

\newcommand\nn{\nonumber}

\newcommand{\evalat}[1]{\,\bigg|_{#1}}

\newcommand{\eqnlab}[1]{\label{eqn:#1}}
\newcommand{\figlab}[1]{\label{fig:#1}}
\newcommand{\tablab}[1]{\label{tab:#1}}
\newcommand{\eqnref}[1]{(\ref{eqn:#1})}

\newcommand{\Eqnref}[1]{Eq.~(\ref{eqn:#1})}
\newcommand{\Figref}[1]{Fig.~\ref{fig:#1}}
\newcommand{\Tabref}[1]{Table~\ref{tab:#1}}
\newcommand{\Eqsref}[1]{Eqs.~(\ref{eqn:#1})}

\newcommand{\applab}[1]{\label{app:#1}}

\newcommand{\seclab}[1]{\label{sec:#1}}

\newcommand{\Appref}[1]{Appendix~\ref{app:#1}}

\newcommand{\Secref}[1]{Section~\ref{sec:#1}}

\newcommand{\Reys}{\ensuremath{\textrm{Re}_s}}
\newcommand{\St}{\ensuremath{\textrm{St}}}

\newcommand{\ar}{\lambda}

\providecommand\bcdot{\boldsymbol{\cdot}}

\newcommand{\supstokes}{{\mathrm{(P)}}}
\newcommand{\supunsteady}{{\mathrm{(U)}}}
\newcommand{\supconvective}{{\mathrm{(C)}}}

\newcommand{\asympt}{\thicksim}

\newcommand{\mma}{Mathematica}

\begin{document}
\title{Rotation of a spheroid in a simple shear at small Reynolds number}
\author{J. Einarsson}
\affiliation{Department of Physics, Gothenburg University, 41296 Gothenburg, Sweden}
\author{F. Candelier}
\affiliation{University of Aix-Marseille, CNRS, IUSTI UMR 7343, 
13 013 Marseille, Cedex 13, France}
\author{F. Lundell}
\affiliation{KTH Royal Institute of Technology, SE-100 44 Stockholm, Sweden}
\author{J. R. Angilella}
\affiliation{Department of Mathematics and Mechanics, LUSAC-ESIX, University of Caen, France}
\author{B. Mehlig}
\affiliation{Department of Physics, Gothenburg University, 41296 Gothenburg, Sweden}

\begin{abstract}
We derive an effective equation of motion for the orientational dynamics of a neutrally buoyant spheroid suspended in a simple shear flow, valid for arbitrary particle aspect ratios and to linear order in the shear Reynolds number.  We show how inertial effects lift the degeneracy of the Jeffery orbits and determine the stabilities of the log-rolling and tumbling orbits at infinitesimal shear Reynolds numbers. For prolate spheroids we find stable tumbling in the shear plane, log-rolling is unstable. For oblate particles, by contrast, log-rolling is stable and tumbling is unstable provided that the aspect ratio is larger than a critical value.  When the aspect ratio is smaller than this value tumbling turns stable, and an unstable limit cycle is born. 
\end{abstract}
% 83.10.Pp Particle dynamics (Fundamentals and theoretical)
% 47.15.G- Low-Reynolds-number (creeping) flows
% 47.55.Kf Particle-laden flows
% 47.10.-g General theory in fluid dynamics

\maketitle

\section{Introduction}
\label{sec:intro}
In this article we describe the effect of weak inertia upon the orientational dynamics of a neutrally buoyant spheroid in a simple shear flow using perturbation theory.
In the absence of inertial effects the rotation of a neutrally buoyant spheroid in a simple shear was determined by Jeffery who found that there are infinitely many degenerate periodic orbits\cite{jeffery1922}, the so-called \lq Jeffery orbits\rq{}. In this limit the initial orientation determines in which way the particle rotates. Fluid and particle inertia lift this degeneracy, but little is known about how this comes about. A notable exception is the work by Subramanian and Koch who have solved the problem for rod-shaped particles in the slender-body approximation \cite{subramanian2005}. We discuss other theoretical results below in Section \ref{sec:background}. 

The question is currently of great interest: several recent papers have reported results of direct numerical simulations (DNS) of the problem, using \lq lattice Boltzmann\rq{} methods \cite{qi2003,huang2012,rosen2014,mao2014}.
These studies reveal that fluid and particle inertia affect the orientational dynamics of a neutrally buoyant spheroid in a simple shear in intricate ways. The DNS are performed at moderate and large shear 
Reynolds numbers, defined as $\Reys=sa^2/\nu$ where $a$ is the largest particle dimension, $s$ is the shear strength and $\nu$ the kinematic viscosity of the suspending fluid.
DNS at very small Reynolds numbers are difficult to perform. But this limit ($\Reys$  of order unity and smaller) is of particular interest. There is a long-standing question whether or not 
a nearly spherical prolate spheroid exhibits stable \lq log-rolling\rq{} 
in this limit, so that its symmetry axis aligns with the vorticity axis. It was first suggested by Saffman that this is the case \cite{saffman1956}, in an attempt to explain Jeffery's hypothesis\cite{jeffery1922} that spheroids rotate in orbits 
that minimise energy dissipation. But stable log-rolling of prolate spheroids
has not been found in DNS, and it 
has been suggested that higher $\Reys$-corrections may explain this discrepancy\cite{mao2014}. The small-$\Reys$ limit is of interest also because it provides stringent tests for DNS. 
\begin{figure}
\includegraphics{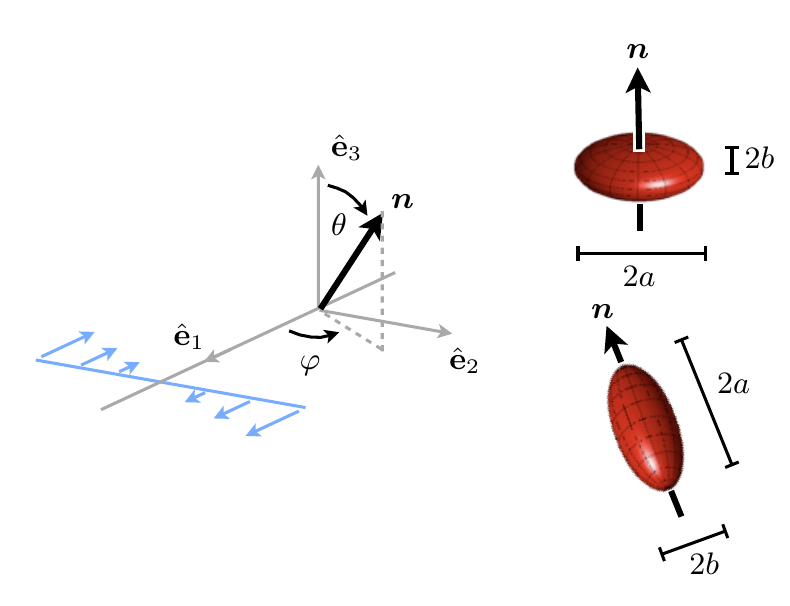}
\caption{\label{fig:1} {\em Color online.} Spheroid rotating in a simple shear (schematic). Shows the Cartesian coordinate system $\hat{\bf e}_j$, $j=1,\ldots,3$.
Vorticity points in the negative $\hat{\ve e}_3$-direction. The flow-shear plane is spanned by $\hat{\bf e}_1$ and $\hat{\bf e}_2$.
The unit vector $\ve n$ points along the symmetry axis of the spheroid. Its polar angle is denoted by $\theta$, the azimuthal angle by $\varphi$.
The major axis length of the spheroid is denoted by $a$, the minor length by $b$.
For prolate particles the aspect ratio is defined as $\lambda = a/b$, and for oblate particles as $\lambda = b/a$.}
\end{figure}
These reasons motivated us to derive an equation of motion that takes into account the effect of weak fluid and particle inertia. 
Our main result is   an approximate dynamical equation for the rotation of a neutrally buoyant spheroid suspended in a simple shear flow, valid 
for arbitrary aspect ratios and to first order in $\Reys$ (Eq.~(\ref{eqn:effective_spherical}) in Section \ref{sec:results}).
In the slender-body limit this equation is of the same form as the one derived in Ref.~\citenum{subramanian2005}. In the completely inertia-free case our results
reduce to Jeffery's equation\cite{jeffery1922}. We find that corrections to this limit arise from both particle inertia (centrifugal and gyroscopic forces), as well as from fluid inertia 
(modifying the hydrodynamic torque on the particle). The particle-inertia corrections we report here 
are consistent 
with earlier numerical and analytical results\cite{lundell2010,einarsson2014}.

Fluid-inertia corrections are taken into account to first order in $\Reys$ using a reciprocal theorem\cite{kim1991}. Our approach is similar to the one adopted in Ref.~\citenum{subramanian2005} in the slender-body limit, but our equation of motion is valid for spheroids with arbitrary aspect ratios. 
By linear stability analysis we determine the stabilities of the periodic orbits of this equation at
infinitesimal $\Reys$ as a function of the particle aspect ratio. The stability calculation 
details how the degeneracy of the Jeffery orbits for a neutrally buoyant spheroid in a simple shear is lifted by weak inertia. 

We find that the log-rolling orbit 
is unstable for prolate particles.
This explains why stable log-rolling is not observed in DNS\cite{qi2003,huang2012,rosen2014,mao2014}
at the smallest shear Reynolds numbers accessible in the simulations.
Moreover we find that tumbling in the flow-shear plane is stable for prolate particles. As the aspect ratio tends to unity there is a bifurcation: for nearly spherical oblate particles log-rolling is stable and tumbling in the flow-shear plane is unstable. 
There is a second bifurcation for oblate particles. At a critical aspect ratio $\lambda_{\rm c} \approx 1/7.3$ tumbling becomes stable and an unstable limit cycle is born. This means that the behaviour of a very flat disk 
depends on its initial orientation for $\lambda < \lambda_{\rm c}$. We discuss how the shape of the limit cycle changes
as the aspect ratio tends to zero.

The remainder of this article is organised as follows. In Section \ref{sec:background} we give an overview over the background of the problem. Section \ref{sec:method} summarises the method employed  in this article, based on a reciprocal theorem \cite{kim1991}. We demonstrate how to calculate the effect of particle and fluid inertia to first order, and how we use the symmetries of the problem to make it tractable. Section \ref{sec:results} summarises our results: the equation of motion and its stability analysis.  We discuss the results in Section \ref{sec:discussion} and conclude with Section \ref{sec:conclusions}.

A brief account of the main results described in this article was given in Ref.~\citenum{einarsson2015}. 
Here  we describe the complete derivation. 
We also present additional results and discussion that could
not be included in the shorter format: we quote precise asymptotic
formulae for small and large aspect ratios, as well as for aspect ratios
close to unity.  We also characterise 
the limit cycle that arises for $\lambda < \lambda_{\rm c}$, 
and compute its linear stability.

\section{Background}
\label{sec:background}
The question of describing the rotation of a neutrally buoyant particle in a simple shear flow has a long history.
Jeffery derived an expression for the torque on an ellipsoidal (tri-axial) particle neglecting inertial effects \cite{jeffery1922}. 
To obtain an equation of motion for small particles he 
assumed that the dynamics is overdamped, that 
the particle rotates so as to instantaneously achieve zero torque. 
This gives rise to Jeffery's equation that is commonly quoted for the special case of spheroidal (axisymmetric) particles. From this equation it follows that spheroids suspended in a simple shear tumble, they stay aligned with the flow direction for some time and then switch orientation by $180$ degrees. The dynamics is degenerate in that there are infinitely many different periodic orbits, the so-called
\lq Jeffery orbits\rq{}. The initial orientation determines which particular orbit is selected. The goal of Jeffery's calculation was to compute the viscosity of a dilute suspension of spheroids, and Jeffery hypothesised that the particles select orbits that minimise energy dissipation. 

Saffman\cite{saffman1956} pointed out that inertial effects lift the degeneracy of the Jeffery orbits, 
and he described  the orientational dynamics of a nearly spherical particle in a simple shear taking into account fluid inertia. For prolate particles he concluded that log-rolling is stable, that tumbling in the shear plane is unstable, and that the stabilities are reversed for oblate particles.
These results are stated in terms of an effective drift for the particle orientation (towards the vorticity axis for prolate particles).
This conclusion supports Jeffery's hypothesis. Saffman did not take into account particle inertia. His method of calculation rests on a joint expansion in small eccentricity and $\Reys$. 

Harper \& Chang\cite{harper1968} addressed the problem in a different way, modeling the dynamics of a rod in a simple shear in terms a dumb-bell, that is two spheres connected by an invisible rigid rod. 
The spheres are subject to Stokes drag and hydrodynamic lift forces\cite{saffman1965}. 
This approximation neglects hydrodynamic interactions between the two spheres, as well as the unsteady term 
in the Navier-Stokes equations. Harper \& Chang arrive at the opposite conclusion, namely that
log-rolling is unstable. Since their result pertains to the slender-body limit the question 
is how the stability of the log-rolling orbit depends on the particle aspect ratio. 

It was subsequently shown by Hinch \& Leal\cite{hinch1972} how weak rotational diffusion breaks the degeneracy of the Jeffery orbits, and their results form the basis for a large part of the work during the last decades on the rheology of dilute suspensions, 
see Refs.~\citenum{petrie1999} and \citenum{lundell2011} for reviews.

Recently there has been a surge of interest in determining the effect of 
weak inertia upon a spheroid tumbling in a simple shear flow in the absence of rotational diffusion.
Subramanian \& Koch\cite{subramanian2005} derived an effective equation of motion
for a  neutrally buoyant rod in the  slender-body limit to first order in fluid and particle inertia.
Their calculation uses a reciprocal theorem\cite{kim1991} and takes into account the unsteady term in the Navier Stokes equation as well as particle inertia. The authors arrive at qualitatively the same conclusion as Harper \& Chang, 
namely that the orientation of the rod eventually drifts towards the flow-shear plane.

In a second paper Subramanian \& Koch\cite{subramanian2006} repeated Saffman's calculation for a neutrally buoyant 
nearly spherical particle. They used a different method, 
similar to the one used in Ref.~\citenum{subramanian2005}, and 
come to the same conclusion as Saffman, that log-rolling is stable
for nearly-spherical prolate particles. 

Recent DNS\cite{qi2003,huang2012,rosen2014,mao2014} have explored the stability
of log-rolling and tumbling orbits, mostly at moderate and large Reynolds numbers, and only for a small
number of aspect ratios. The simulations show unstable log-rolling for prolate particles 
at the smallest Reynolds numbers studied.
We note that You, Phan-Thien \& Tanner\cite{you2007} misquote Saffman when they describe their numerical results on the rotation of a spheroid in a Couette flow at Reynolds numbers of the order of $10$ and larger. In the introduction of Ref.~\citenum{you2007} it is implied that Saffman's theory\cite{saffman1956} predicts that nearly spherical prolate particles tend to the flow-shear
plane.

\section{Method}
\label{sec:method}
In this section  we give a brief but complete summary of our calculation. The most technical details and tabulations are deferred to appendices.  We start with notation and the relevant dimensionless parameters 
determining the physics. Then we give the governing equations and explain how to express the hydrodynamic torque through a reciprocal theorem\cite{lorentz,happel1983,kim1991,subramanian2005}. Finally we explain the perturbation scheme and list the symmetries that  
severely constrain the form of the solution.

\subsection{Notation}
The calculations described in this paper involve vectors and tensors in three spatial dimensions. We employ index notation with the implicit summation convention for repated indices, and we use the  Kronecker ($\delta_{ij}$) 
and Levi-Civita ($\lc_{ijk}$) tensors. 

\subsection{Units and dimensionless numbers}\seclab{units}
The physics of the problem is governed by three dimensionless numbers: the shear Reynolds number $\Reys$ (measuring fluid inertia), the Stokes number $\St$ (measuring particle inertia) and the particle aspect ratio $\lambda$. 

We work with dimensionless variables. The length scale is given by  the particle major axis $a$. The velocity scale is taken to be $sa$ where $s$ is the shear rate. The explicit time dependence of the flow 
(the time scale for the unsteady fluid inertia) scales
as $\thicksim 1/s$ since it is determined by
the particle angular velocity because to lowest order the unsteadiness arises from the particle motion.
The corresponding scale for pressure is $\mu s$, and  force and torque are measured in units of  $\mu s a^2$ and $\mu s a^3$, respectively.

From these scales the dimensionless parameters are formed. As mentioned in the Introduction,
the shear Reynolds number is defined as
\begin{align}
  \Reys = \frac{sa^2\rho_{\rm f}}{\mu}\,.
\end{align}
Here $\rho_{\rm f}$ is the density and $\mu=\rho_{\rm f}\nu$ 
is the dynamic viscosity of the surrounding fluid ($\nu$ is the kinematic viscosity).

The  Stokes number, measuring the particle inertia, is given by the ratio of the typical rate of change of angular momentum and the typical torque:
\begin{align}
  \St 
  = \frac{\rho_{\rm p} sa^2}{\mu} = \frac{\rho_{\rm p}}{\rho_{\rm f}}\Reys.
\end{align}
Here $\rho_{\rm p}$ is the particle density.
For a neutrally buoyant particle, $\rho_{\rm p} = \rho_{\rm f}$, we have $\St=\Reys$.

We define the particle aspect ratio $\lambda$ as the ratio between the length along the symmetry axis and the length 
transverse to the symmetry axis (Fig. \ref{fig:1}). That is $\lambda>1$ denotes prolate particles while $\lambda<1$ denotes oblate particles. Because we measure length in units of the major particle axis $a$, the aspect ratio $\lambda$ of a prolate particle is $a/b$, while the aspect ratio of an oblate particle is $b/a$, where $b$ denotes the length of the minor axis of the particle
(see Fig.~\ref{fig:1}).

\subsection{Equations of motion}
 Let $n_i$ denote the components of the unit vector $\ve n$ pointing in the direction of the particle symmetry axis (Fig. \ref{fig:1}), and $\omega_i$ the components of the angular velocity of the particle. 
Newton's second law for the orientational degrees of freedom for an axisymmetric particle reads:
\begin{align}
  &{\phantom\St}\dot n_i = \lc_{ijk}\omega_jn_k\,,\quad
  \St \left[\dot I_{ij}\omega_j+ I_{ij}\dot \omega_j\right] =T_i\,.\eqnlab{eqnofmotion}
\end{align}
Dots denote time derivatives, and $I_{ij}$ are the elements of the moment-of-inertia tensor of the particle, and $T_i$ is the torque exerted on the particle.
The moment-of-inertia tensor of an axisymmetric particle with axis of symmetry $\ve n$ is on the form
\begin{align}
  I_{ij} &= A^I n_in_j + B^I(\delta_{ij}-n_in_j)\,,\eqnlab{momentofinertia}
\end{align}
where $A^I$ and $B^I$ correspond to the moments-of-inertia around and transverse to the symmetry axis.
Using the dimensionless variables  introduced in \Secref{units} we have for a prolate spheroid ($\lambda>1$)
\begin{align}
  A^I = \frac{8\pi}{15}\frac{1}{\lambda^4},\qquad \qquad B^I=\frac{4\pi}{15}\frac{1}{\lambda^2}\left(1+\frac{1}{\lambda^2}\right)
\end{align}
and for an oblate spheroid ($\lambda<1$)
\begin{align}
  A^I = \frac{8\pi}{15}\lambda,\qquad \qquad B^I=\frac{4\pi}{15}\lambda\left(1+\lambda^2\right).
\end{align}
We rewrite the equation of motion \eqnref{eqnofmotion} as
\begin{align}
  \dot{\omega}_i &= - I^{-1}_{ij}\dot{I}_{jk}\omega_k + \frac{1}{\St}I^{-1}_{ij}T_j 
  = -\frac{A^I-B^I}{B^I}\lc_{ijk}\omega_jn_k n_l\omega_l + \frac{1}{\St}I^{-1}_{ij}T_j \,. \eqnlab{eqnofmotion2}
\end{align}
In the final step we used the definition \eqnref{momentofinertia} of $I_{ij}$ and the equation of motion \eqnref{eqnofmotion} for $\dot{n}_i$. 

In this paper, the  $T_i$ are the elements of  the hydrodynamic torque exerted on the particle by the fluid. In \Secref{torque} we formulate the hydrodynamic torque to $O(\Reys)$ via the reciprocal theorem. In \Secref{angularvelocity} we perturbatively compute the resulting angular velocity to order $O(\Reys)$ and $O(\St)$.

\subsection{Calculation of the hydrodynamic torque to order {\protect\textnormal{$\Reys$}}}
\seclab{torque}
The straightforward approach to determine the torque on a particle in a fluid is to solve Navier-Stokes equations for the velocity and pressure fields, then compute the stress tensor and finally integrate the stress tensor over the surface of the particle. The reciprocal theorem\cite{lorentz,kim1991,subramanian2005} offers an alternative, and often more convenient, route to the hydrodynamic forces. In particular, we may avoid solving for the complete flow field.
In this Section we specify the Navier-Stokes problem we need to solve, and explain how we use the reciprocal theorem to simplify the calculations.

{\em Navier-Stokes problem for the disturbance flow.}
We consider a particle with boundary $\mathcal S$ immersed in an linear ambient flow $(\ve u^\infty, p^\infty)$. 
Throughout this paper we express the ambient flow as
\begin{align}
u_i^\infty&=A_{ij}^\infty r_j = \lc_{ikj} \Omega_k^\infty r_j + S_{ij}^\infty r_j,
\end{align}
or equivalently with $\lc_{ikj} \Omega_k^\infty =  O_{ij}^\infty$
\begin{align}
u_i^\infty&= O_{ij}^\infty r_j+ S_{ij}^\infty r_j.
\end{align}
Here $\ma S^\infty$ and $\ma O^\infty$ are the symmetric and antisymmetric parts of the flow gradient, given by
\begin{align}
S^\infty_{ij} &= \frac{1}{2}\left(A_{ij}^\infty \!+\! A_{ji}^\infty\right)\,,\qquad
O^\infty_{ij} = \frac{1}{2}\left(A_{ij}^\infty \!-\! A_{ji}^\infty\right)\,.
\end{align}
In dimensionless variables (\Secref{units}) the Navier-Stokes equations read
\begin{align}
  \Reys \left(\partial_t u_i + u_j \partial_j u_i\right) &= -\partial_i p + \partial_j \partial_j u_i\,, \,\,\, \partial_i u_i &= 0\,.
\end{align}
Note that the unsteady and convective inertia terms come with the same prefactor in this problem. This happens because the timescale of the particle motion is the same as the timescale of the flow.
The boundary condition is no-slip on the surface of the particle
\begin{align}
  u_i &\!=\! \lc_{ijk}\omega_jr_k\,\, {\rm for}\,\,\ve r \!\in\! S\,, \quad
  u_i \!= \!u_i^\infty\,\,\,\mbox{as}\,\,
  |\ve r|\!\to\!\infty\,.
\end{align}
We introduce the disturbance field $(u'_i, p')$ from the particle 
\begin{align}
u_i &= u^\infty_i + u'_i\,,\quad
  p = p^\infty + p'\,.
\end{align}
If we assume that $(\ve u^\infty, p^\infty)$ satisfies the 
Navier-Stokes equations we have the disturbance problem
\begin{align}
\Reys\left(\partial_t u'_i + u^\infty_j \partial_j u'_i + u'_j \partial_j u^\infty_i + u'_j \partial_j u'_i\right)
= -\partial_i p' + \partial_j \partial_j u'_i\,, \eqnlab{nsdisturbance}
\end{align}
and the boundary conditions is expressed in the slip angular velocity $\Omega_i = \Omega^\infty_i-\omega_i$ as
\begin{align}
u'_i &= -\lc_{ijk}\Omega_j r_k - S_{ij}^\infty r_j, \,\, \ve r \in S, \nn\\
  u'_i &= 0\,, \qquad  |\ve r|\to\infty\,.\eqnlab{nsdisturbancebc}
\end{align}
Finally, when applying the reciprocal theorem we shall use that, by definition, the divergence of the stress tensor satisfies the following equalities:
\begin{align}
   \partial_j \sigma'_{ij} &= -\partial_i p' + \partial_j \partial_j u'_i \nn \\
   &= \Reys \left(\partial_t u'_i + u^\infty_j \partial_j u'_i + u'_j \partial_j u^\infty_i + u'_j \partial_j u'_i\right) \nn \\
   &\equiv \Reys f_i (\ve u')\,.\eqnlab{fdef}
\end{align}

{\em The Stokes solution.}
This paper concerns a spheroidal particle suspended in a linear flow. We thus need explicit solutions to \Eqnref{nsdisturbance} at $\Reys=0$ in this geometry. We use a finite multipole expansion\cite{chwang1975,kim1991} (see \Appref{stokessolutions}). In our notation they read
\begin{align}
  u'_i &= \mathcal Q^R_{ij,k}\varepsilon_{jkl}\left[
    \left(A^R n_ln_m + B^R(\delta_{lm} - n_ln_m)\right)\Omega_m
   + C^R \varepsilon_{lmn} n_m S^\infty_{no}n_o\right] \nn\\
  & + \left(\mathcal Q^S_{ij,k} + \alpha \mathcal Q^Q_{ij,llk} \right) \eqnlab{stokessol}\\
  &\times\left[\left(A^S n^A_{jklm} + B^S n^B_{jklm} + C^S n^C_{jklm}\right)S^\infty_{lm} 
     - C^R \left(\varepsilon_{jlm}n_kn_m + \varepsilon_{klm}n_jn_m\right)\Omega_l
  \right]\,,\nn
\end{align}
where
\begin{align*}
  n^A_{jklm} &= (n_jn_k - \frac{1}{3}\delta_{jk})(n_ln_m - \frac{1}{3}\delta_{lm}), \\
  n^B_{jklm} &= n_j\delta_{kl}n_m + n_k\delta_{jl}n_m + n_j\delta_{km}n_l + n_k\delta_{jm}n_l 
  - 4n_jn_kn_ln_m, \\
  n^C_{jklm} &= -\delta_{jk}\delta_{lm} + \delta_{jl}\delta_{km} + \delta_{kl}\delta_{jm} \\
        &\qquad+ \delta_{jk} n_ln_m + \delta_{lm}  n_jn_k 
          - n_j\delta_{kl}n_m - n_k\delta_{jl}n_m\\
          &\qquad  - n_j\delta_{km}n_l- n_k\delta_{jm}n_l + n_jn_kn_ln_m.
\end{align*}
Here $A^R$, $B^R$, $C^R$, $A^S$, $B^S$, $C^S$ and $\alpha$ are known constants 
that depend on the particle aspect ratio $\lambda$. The exact definition of the spheroidal multipoles $\mathcal Q$ and the values of all constants are given in \Appref{stokessolutions}, see in particular \Tabref{ARBRCR}.

{\em The reciprocal theorem.}
This theorem\cite{lorentz,happel1983,kim1991,subramanian2005} relates integrals of the velocity and stress fields of two incompressible and 
Newtonian fluids. The idea is the following. Let one set of fields represent the actual problem of interest, the \emph{primary problem}. Then choose the second set of fields to be an \emph{auxiliary problem} with known solution, such that an integral in the theorem relates to hydrodynamic torque of the primary problem. Provided that all integrals in the theorem converge and can be evaluated, we can solve the resulting equations for the hydrodynamic torque.

The reciprocal theorem for the two sets $(\tilde u_i, \tilde \sigma_{ij})$ and $(u'_i, \sigma'_{ij})$ can be stated as
\begin{align}
  &\int_S \rd \tilde F_i u'_i + \int_V \rd V u'_i \partial_j \tilde \sigma_{ij}
= \int_S \rd F'_i \tilde u_i + \int_V \rd V \tilde u_i \partial_j \sigma'_{ij}\,.
\end{align}
Here $\rd F_i = \sigma_{ij}\xi_j\rd S$ is the differential force from the fluid on the surface element with normal vector $\xi_j\rd S$. The volume integrals are to be taken over the entire fluid volume outside the particle, and the surface integrals over all surfaces bounding the fluid volume, with surface normals pointing out of the fluid volume.

In the following we apply the reciprocal theorem to the calculation of the hydrodynamic torque on a particle.

 {\em Calculation of the torque.}
We choose the auxiliary problem $(\tilde u_i, \tilde \sigma_{ij})$ to be the Stokes flow around an identical particle rotating with an angular velocity $\tilde\omega_i$ in an otherwise quiescent fluid. Its solution is given by \Eqnref{stokessol} with $\ve u^\infty=0$. The primary problem is the disturbance problem defined in \Eqnref{nsdisturbance}. Inserting the boundary conditions into the reciprocal theorem yields
\begin{align}
& \int_{\mathcal S} \rd \tilde F_i (\lc_{ijk}(\omega_j-\Omega^\infty_j)r_k - S^\infty_{ij}r_j) 
= \int_{\mathcal S} \rd F'_i \lc_{ijk}\tilde\omega_jr_k  + \Reys \int_V \rd V \tilde u_i f_i(\ve u')\,.\eqnlab{rt_2}
\end{align}
We also used that $\partial_j\tilde\sigma_{ij}=0$.
This equality holds because $\tilde u_i$ is a Stokes flow.
Both primary and auxiliary velocity fields vanish as $|\ve r|\to\infty$, therefore both integration surfaces are only the particle surface $\mathcal S$. Note that the surface integrals are to be taken with surface normals out of the fluid domain, so that $\rd F_i$ is the differential force exerted on the particle by the fluid.
In the integrals we identify the hydrodynamic torque on the particle, it is given by
\begin{align}
  T_j &= \int_{\mathcal S} \rd F_i \lc_{ijk} r_k\,.
\end{align}
It follows:
\begin{align}
 & (\omega_j-\Omega^\infty_j)\tilde T_j - \int_{\mathcal S} \rd \tilde F_i S^\infty_{ij}r_j 
= \tilde\omega_j (T_j-T^\infty_j)  + \Reys\int_V \rd V \tilde u_i f_i(\ve u')\,.
\label{eqn:rc3}
\end{align}
The auxiliary torque $\tilde T_j$ together with the surface integral add up to the Jeffery torque\cite{jeffery1922} $T_j^{(0)}$: 
\begin{align}
        T^{(0)}_j  &= c_\xi\left[ \left(A^R n_jn_k + B^R(\delta_{jk}-n_jn_k)\right)(\Omega^\infty_k-\omega_k) 
+  C^R \lc_{jkm}n_kn_l  S^\infty_{ml}\right]\,.
\end{align}
The constant $c_\xi$ is given in \Tabref{ARBRCR} in \Appref{stokessolutions}. 
The contribution 
\begin{align}
  T^\infty_j &\equiv \int_{\mathcal S} \rd S\sigma^\infty_{il}\xi_l \lc_{ijk}\tilde\omega_jr_k 
\end{align}
evaluates to zero for any linear flow $u_i^\infty$.
It follows that \Eqnref{rc3} becomes
\begin{align}
 \tilde\omega_j T_j 
 &=  \tilde \omega_j T^{(0)}_j - \Reys\int_V \rd V \tilde u_i f_i(\ve u')\,.
\end{align}
Since $\tilde u_i$ is linear in $\tilde \omega_j$, this variable can
be eliminated. We finally obtain:
\begin{align}
  T_j  &=   T^{(0)}_j - \Reys\int_V \rd V \tilde U_{ij} f_i(\ve u')\,,
\eqnlab{rt3}
\end{align}
where
\begin{align}
    \tilde U_{ip} &= -\mathcal Q^R_{ij,k}\varepsilon_{jkl}\left[
    \left(A^R n_ln_p + B^R(\delta_{lp} - n_ln_p)\right) 
  \right] \\
  & \quad \hspace*{-5mm}+ \left(\mathcal Q^S_{ij,k} + \alpha \mathcal Q^Q_{ij,llk} \right)\left[
     C^R \left(\varepsilon_{jpm}n_kn_m + \varepsilon_{kpm}n_jn_m\right)
  \right]\,.\nn
\end{align}
Thus far we have made no approximation, and \Eqnref{rt3} is exact, 
the difficulty lies in evaluating the Navier-Stokes disturbance flow $\ve u'$. 
This is a complicated non-linear problem since  $T^{(0)}_j$, $\tilde U_{ij}$ and $f_i(\ve u')$ all depend on the direction $\ve n$ and upon the angular velocity $\ve \omega$ of the particle. The flow equations thus couple non-linearly to the rigid body equations of motion for the particle. In the following we solve this system of equations in perturbation theory valid to first order in $\St$ and $\Reys$.

\subsection{Perturbative calculation of the particle angular velocity}\seclab{angularvelocity}
In this section we determine the angular velocity $\ve \omega$ of the particle to lowest order in $\St$ and $\Reys$, assuming that both $\St$ and $\Reys$ are small, so that $\Reys\St$ is negligible.
We recall the equation of motion \eqnref{eqnofmotion2} for the particle orientation, 
and insert the expression for the hydrodynamic torque obtained in \Secref{torque}:
\begin{align}
  \dot{n}_i &= \lc_{ijk}\omega_jn_k\,, \nn \\
  \St \,\dot{\omega}_i &= - \St \frac{A^I-B^I}{B^I}\lc_{ijk}\omega_jn_k n_l\omega_l + I^{-1}_{ij}T^{(0)}_j
- \Reys I^{-1}_{ij}\int_V \rd V \tilde U_{kj} f_k(\ve u')  \,. \eqnlab{eqnofmotion3}
\end{align}
Now we expand the angular velocity as
\begin{align}
  \omega_i &= \omega^{(0)}_i + \St \omega_i^{(\St)} + \Reys \omega_i^{(\Reys)} + o(\Reys,\St). \eqnlab{omegaansatz}
\end{align}
Next, we insert these expansions into the equation of motion \eqnref{eqnofmotion2} and collect terms of equal order in $\St$ and $\Reys$:
\begin{align}
0&= T_j^{(0)}\,,\nn\\
\dot \omega^{(0)}_i &=  -\frac{A^I-B^I}{B^I}\lc_{ijk}\omega^{(0)}_jn_k n_l\omega^{(0)}_l 
- I^{-1}_{ij}c_\xi \left(A^R n_jn_k + B^R(\delta_{jk}-n_jn_k)\right)\omega_i^{(\St)}\,, \nn\\
0&= c_\xi \left(A^R n_jn_k + B^R(\delta_{jk}-n_jn_k)\right) \omega_i^{(\Reys)}
 + \int_V \rd V \tilde U_{kj} f_k(\ve u')\,.
\label{eq:29}
\end{align}
In the last term it is understood that the volume integral need only be evaluated to $O(1)$, so that we may use the Stokes flow solutions for $\ve u'$.
The first equation gives the Jeffery angular velocity $\omega_i^{(0)}$,
\begin{align}
  \omega^{(0)}_i &= \Omega^\infty_i + \frac{C^R}{B^R}\lc_{ikm}n_kn_l  S^\infty_{ml}\,. \eqnlab{omega0}
\end{align}
The dynamics of $\ve n$ is to lowest order given by
\begin{align}
  \dot{n}_i^{(0)} &= \lc_{ipq}\omega^{(0)}_{p}n_q 
 \lc_{ipq}\Omega^\infty_{p}n_q + \frac{C^R}{B^R}\left(S^\infty_{ip}n_p - n_in_pn_qS^\infty_{pq}\right)\,.\eqnlab{njeffery}
\end{align}
From \Tabref{ARBRCR} in \Appref{stokessolutions} we infer that for both prolate and oblate spheroids
\begin{align}
  \frac{C^R}{B^R} &= \Lambda = \frac{\lambda^2-1}{\lambda^2+1}\,.
\end{align}
This shows that Eq.~(\ref{eqn:njeffery}) is Jeffery's equation\cite{jeffery1922} for the orientational dynamics of a spheroid in a simple shear.

The two remaining equations in (\ref{eq:29}) may be inverted to
\begin{align}
  \omega_i^{(\St)} &= 
 \frac{1}{c_\xi} \left(\frac{1}{A^R} n_in_j + \frac{1}{B^R}(\delta_{ij}-n_in_j)\right)\nn\\
 &\quad \times \left[-I_{jk}\dot \omega^{(0)}_k- \frac{A^I-B^I}{B^I}I_{jk}\lc_{klm}\omega^{(0)}_ln_m n_p\omega^{(0)}_p\right]\nn\\
 \omega_i^{(\Reys)}&= -\frac{1}{c_\xi} \left(\frac{1}{A^R} n_in_j + \frac{1}{B^R}(\delta_{ij}-n_in_j)\right)
 \int_V \rd V \tilde U_{kj} f_k(\ve u').\eqnlab{omega1}
\end{align}
\Eqnref{omegaansatz} together with \Eqsref{omega0} and (\ref{eqn:omega1}) yield the effective angular velocity under the effect of weak particle and fluid inertia. From the equation of motion \eqnref{eqnofmotion2} we define the effective vector field 
\begin{align}
  \dot n_i &= \lc_{ijk}\omega_j n_k \equiv \dot n_i^{(0)}+\St\, \dot n_i^{(\St)}+\Reys\, \dot n_i^{(\Reys)}.\eqnlab{ndoteffective}
\end{align}
This vector field describes the time evolution of $\ve n$. The first term is the Jeffery vector field (\ref{eqn:njeffery}). The two new terms represent the effects of particle inertia and fluid inertia. The terms due to particle inertia are straightforward to evaluate directly, but the volume integral in \Eqnref{omega1} is very tedious to evaluate. To make the calculation feasible we
exploit the  symmetries of the problem.

\subsection{Symmetries of the effective equation of motion}
Both correction terms in \Eqnref{ndoteffective} are quadratic in the ambient flow gradient tensor $A^\infty_{ij}$. In other words, they are on the form
\begin{align}
  \dot n_i &\!=\! C^{(1)}_{ijklm}O^\infty_{jk}O^\infty_{lm} \!+\! C^{(2)}_{ijklm}O^\infty_{jk}S^\infty_{lm} \!+\! C^{(3)}_{ijklm}S^\infty_{jk}S^\infty_{lm},\eqnlab{nsymmansatz}
\end{align}
where the tensorial coefficients $C^{(i)}_{ijklm}$ are composed of the remaining available tensor quantities: $n_p$ and $\delta_{pq}$ ($\lc_{ijk}$ is already used in $O^\infty_{ij}$). We make an exhaustive enumeration of all possible combinations, and then use the symmetries listed in \Tabref{symm} to remove or combine items in the list. For example we start by letting
\begin{align}
&  C^{(1)}_{ijklm} = \sum_{\mathrm{P}}\left[\eta_1^{(\mathrm{P})}n_{p_1} \delta_{p_2p_3}\delta_{p_4p_5}
+\eta_2^{(\mathrm{P})} n_{p_1} n_{p_2}n_{p_3}\delta_{p_4p_5}
+\eta_3^{(\mathrm{P})} n_{p_1} n_{p_2}n_{p_3} n_{p_4}n_{p_5}\right]\eqnlab{csymm}
\end{align}
where the sum is over all $5!\!=\!120$ permutations $\mathrm{P}$ of $(i,j,k,l,m)$, and $\eta_i^{\mbox{\tiny (P)}}$ are unique coefficients for each term. We include only odd powers of $n_i$ as any even terms would break the particle inversion symmetry. We then insert this enumeration into the first term of \Eqnref{nsymmansatz}, and contract and apply the first three symmetries in \Tabref{symm} until we reach a list of unique candidate terms. In this case the only two unique terms turn out to be $O_{ij}O_{jk}n_k$ and $n_i n_jO_{jk}O_{kl}n_l$. Finally, we use the fact that the equation of motion may not change the magnitude of the unit vector $\ve n$. This constraint forces the coefficients of the two unique terms to be the same magnitude but opposite sign. Upon renaming the coefficients $\pm\alpha_1$ we get the first term in \Eqnref{symmalpha}. The other terms are derived similarly by inserting \eqnref{csymm} into the other terms in \Eqnref{nsymmansatz}. The result contains only six independent terms:
\begin{align}
  \dot n_i &= \alpha_1 \left(\delta_{ij}\!-\!n_in_j\right) O^\infty_{jk}O^\infty_{kl}n_l \nn\\
&+ \alpha_2 \left(\delta_{ij}\!-\!n_in_j\right) S^\infty_{jk}O^\infty_{kl}n_l \nn\\
            &+ \alpha_3 \left(\delta_{ij}\!-\!n_in_j\right) O^\infty_{jk}S^\infty_{kl}n_l \nn\\&+ \alpha_4 \left(\delta_{ij}\!-\!n_in_j\right) S^\infty_{jk}S^\infty_{kl}n_l \nn\\
            &+ \alpha_5 (n_pS^\infty_{pq}n_q)O^\infty_{ij}n_j \nn\\
            &+ \alpha_6 (n_pS^\infty_{pq}n_q)\left(\delta_{ij}\!-\!n_in_j\right)S^\infty_{jk}n_k\,.\eqnlab{symmalpha}
\end{align}
Here the scalar functions $\alpha_1,\ldots,\alpha_6$ 
are linear in $\St$ and $\Reys$,
and depend on the aspect ratio $\lambda$ in a non-linear (and unknown) way.
These  coefficients are determined by evaluating the vector field in \Eqnref{ndoteffective} for six independent directions of $\ve n$ and solving the 
resulting system of linear equations for $\alpha_1,\ldots,\alpha_6$.

\begin{table}
\begin{ruledtabular}
\caption{\tablab{symm} Symmetries constraining the form of the effective equation of motion, \Eqnref{ndoteffective}.}
\begin{tabular}{lr}
  $S^\infty_{ii} = 0$ & Incompressible flow  \\
  $S^\infty_{ij} = S^\infty_{ji}$ & $\ma S^\infty$ symmetric  \\
  $O^\infty_{ij} = -O^\infty_{ji}$ & $\ma O^\infty$ anti-symmetric  \\
  $n_i \dot n_i = 0$ & Dynamics preserves magnitude \\
  $n_i \to -n_i \implies \dot n_i \to -\dot n_i$ & Particle inversion symmetry \\
\end{tabular}
\end{ruledtabular}
\end{table}

In the particular case of a simple shear flow, we have explicitly (see \Figref{1} for the geometry)
\begin{align}
 O^\infty_{ij} \!=\! \frac{1}{2}\left(\delta_{i1}\delta_{j2} \!-\! \delta_{i2}\delta_{j1}\right), \, S^\infty_{ij} \!=\! \frac{1}{2}\left(\delta_{i1}\delta_{j2} \!+\! \delta_{i2}\delta_{j1}\right).\eqnlab{sheardef}
\end{align}
We observe that for the simple shear $O^\infty_{ij}O^\infty_{jk}= -S^\infty_{ij}S^\infty_{jk}$, and $S^\infty_{ij}O^\infty_{jk} = -O^\infty_{ij}S^\infty_{jk}$. Then the form of the equation of motion 
simplifies to
\begin{align}
  \dot n_i &= \beta_1 (n_pS^\infty_{pq}n_q)\left(\delta_{ij}-n_in_j\right)S^\infty_{jk}n_k  \nn\\
            &+ \beta_2 (n_pS^\infty_{pq}n_q)O^\infty_{ij}n_j  \nn\\
            &+ \beta_3 \left(\delta_{ij}-n_in_j\right) O^\infty_{jk}S^\infty_{kl}n_l \nn\\
            &+ \beta_4 \left(\delta_{ij}-n_in_j\right) S^\infty_{jk}S^\infty_{kl}n_l\,,\eqnlab{nansatz}
\end{align}
with
\begin{align}
  \beta_1 &= \alpha_6\,, \nn\\ \beta_2 &= \alpha_5\,, \nn\\
  \beta_3 &= \alpha_3 \!-\! \alpha_2\,,\nn\\ \beta_4 &= \alpha_4 \!-\! \alpha_1\,.
\end{align}
Thus, for the case of the simple shear flow it suffices to evaluate the effective vector field in \Eqnref{ndoteffective}, in particular the volume integral in \Eqnref{omega1}, with four independent values of $\ve n$ in order to solve for the unknown scalar coefficients $\beta_1,\ldots,\beta_4$.

\subsection{Evaluation of the volume integral in Eq.~(\ref{eqn:omega1})}
The volume integral in Eq.~(\ref{eqn:omega1}) contains four distinct terms: 
$\partial_t u'_i$, represents unsteady fluid inertia, and 
the three terms
$u^\infty_j \partial_j u'_i + u'_j \partial_j u^\infty_i + u'_j \partial_j u'_i$ 
represent convective fluid inertia.
We compute these four terms using the explicit Stokes-flow solutions \eqnref{stokessol}. While the 
Stokes flow has no explicit time dependence, both particle direction $\ve n$ and angular velocity $\ve \omega$ do. Thus each occurence of $n_k$ and $\omega_k$ has to be differentiated to compute the contribution due to unsteady fluid inertia.
The differentiation and tensor contractions are implemented by a custom set of pattern matching rules in {\mma}\textsuperscript{\textregistered}. The calculation is both long and error prone. 
We have therefore automated every possible step, including solving the Stokes-flow equations.

We demonstrate the remainder of the procedure by a small example. 
Consider the contribution in the $\hat{\bf e}_3$-direction of \Eqnref{omega1}
due to unsteady fluid inertia: 
\begin{align}
 -\delta_{i3}\frac{1}{c_\xi} \left(\frac{1}{A^R} n_in_j + \frac{1}{B^R}(\delta_{ij}-n_in_j)\right)
 \frac{\Reys}{\St}\int_V \rd V \tilde U_{kj} \partial_t u'_k \,.
\end{align}
We first perform the time derivatives on \eqnref{stokessol} in the manner explained above. Then we insert the components of $\ve n$, and the explicit form of the shear flow \eqnref{sheardef}. At this point we can explicitly perform the sum over all repeated indices. The result in this example consists of $858$ terms, after collecting terms with same spatial dependence. The terms have a prefactor that stem from 
the Stokes-flow coefficients (see \Appref{stokessolutions}), and a 
spatial dependence coming from $r_i$ and the spheroidal integrals $J^n_m$ and $K^n_m$ 
(see \Appref{spheroidalintegrals}). 
For $\ve n=[1/2, \sqrt{3}/2,0]$  a typical term looks like this:
\begin{align}
  \frac{1575 \alpha ^2 C^R  (A^S-3 C^S) (2 B^R+C^R) r^2 r_2^3 K^0_7 K^1_9}{16 (B^R)^2 c_\xi} \nn
\end{align}
We note that the only spatial dependence on the azimuthal angle around the symmetry axis of the body comes from factors $r_i$. We introduce a rotated coordinate system in which $r_i = R_{ji}r'_j$, such that $r'_1$ is along the particle symmetry axis (see \Appref{spheroidalintegrals}). This change of basis enables integration of one spatial coordinate.

After this operation $260$ terms still remain which we program {\mma}\textsuperscript{\textregistered}
to express in spheroidal coordinates (\Appref{spheroidalcoords}) and integrate over the remaining two spatial coordinates.

As a consistency check we have also evaluated the volume integral numerically over all three spatial dimensions by converting to spheroidal coordinates and choosing a specific value of $\lambda$. For extreme values of $\lambda$ the numerics are difficult, nevertheless they serve as a check for a wide range of aspect ratios (see markers in \Figref{pof_betas}).

\section{Results} 
\label{sec:results}
\subsection{Effective equation of motion}
We parametrize the vector $\ve n$ in a spherical coordinate system $(\theta, \varphi)$ with $\theta$ the polar angle and $\varphi$ the azimuthal angle
(\Figref{1}):
\begin{align*}
  n_1 &= \sin\theta\cos\varphi\,\quad 
n_2= \sin\theta\sin\varphi\,,\quad
   n_3 = \cos\theta\,.
\end{align*}
In these coordinates \Eqnref{nansatz} is expressed as
\begin{widetext}
\begin{subequations}\eqnlab{effective_spherical}
\begin{align}
  \dot \varphi(\theta, \varphi) &=\frac{1}{2}\left(\Lambda  \cos 2 \varphi -1\right) +  \frac{1}{8} \beta_1 \sin ^2\theta \sin 4 \varphi -\frac{1}{4} \sin 2 \varphi \left(\beta_2 \sin ^2\theta+\beta_3\right)\,,
  \eqnlab{phi}\\
  \dot \theta(\theta, \varphi) &= \Lambda  \sin \theta  \cos \theta \sin \varphi  \cos \varphi + \frac{1}{4} \sin \theta \cos \theta \left(\beta_1 \sin ^2\theta \sin ^2 2\varphi+\beta_3 \cos 2 \varphi +\beta_4\right)\,. \eqnlab{theta}
\end{align}
\end{subequations}
\end{widetext}
We compute the contributions to $\beta_\alpha$ from three sources: particle inertia, unsteady fluid inertia and convective fluid inertia. Although the result is only valid for neutrally buoyant particles ($\Reys=\St$), it is interesting to consider the contributions separately:
\begin{align}
	\beta_\alpha &= \St \beta_\alpha^\supstokes + \Reys \beta_\alpha^\supunsteady + \Reys \beta_\alpha^\supconvective
\end{align}
The contribution from particle inertia is straightforward to compute and can be expressed in closed form as
\begin{align}
  \beta_1^\supstokes &=  \frac{2 B^I (C^R)^2}{(B^R)^3 c_\xi},                  \nn\\
  \beta_2^\supstokes  &= -\frac{C^R (A^I-2 B^I)}{(B^R)^2 c_\xi},               \nn\\
\beta_3^\supstokes &=  \frac{A^I C^R}{(B^R)^2 c_\xi},                        \nn\\ \beta_4^\supstokes &= -\frac{(A^I-B^I) (B^R)^2+B^I (C^R)^2}{(B^R)^3 c_\xi}.   \eqnlab{betas_stokes}
\end{align}
The coefficients on the r.h.s. of these equations
are tabulated for both prolate and oblate spheroids in \Tabref{ARBRCR} in \Appref{stokessolutions}. The coefficients in \Eqnref{betas_stokes} are shown as dotted lines in \Figref{pof_betas}.
\begin{table}[t]
\begin{ruledtabular}
\caption{\tablab{asymptotic} Asymptotic results for $\beta_\alpha$. Contributions from
particle inertia, unsteady fluid inertia, and convective fluid inertia are shown separately. Factors of $\Reys$ and $\St$ are omitted.}
\begin{tabular}{lcccc}
\multicolumn{5}{c}{Thin oblate particles ($\lambda\to0$)}\\
& Total & Unsteady & Convective & Particle \\
$\beta_1$&
$\frac{11}{30}$  &
$\frac{1}{5}$    &
$\frac{1}{6}$&
$0$ \\
$\beta_2$& $\frac{1}{10}$  &    $-\frac{1}{20}$ &       $\frac{3}{20}$&  $0$\\
$\beta_3$& $-\frac{1}{5}$  &    $-\frac{3}{20}$ &       $-\frac{1}{20}$&  $0$\\
$\beta_4$& $-\frac{1}{3}$  &    $-\frac{3}{20}$ &       $-\frac{11}{60}$&  $0$ \\
\noalign{\vskip 1ex} \hline
\multicolumn{5}{c}{Nearly spherical particles ($|\epsilon|\ll1$)}\\
& Total & Unsteady & Convective & Particle \\
$\beta_1$&$\frac{137 \epsilon ^2}{294}$                                                 & $0$                                                   & $\frac{163 \epsilon ^2}{490}$                                 & $\frac{2 \epsilon ^2}{15}$\\
$\beta_2$&$\frac{2\epsilon}{21}+\frac{81 \epsilon ^2}{245}$     & $\frac{62 \epsilon ^2}{525}$  & $\frac{\epsilon}{35}+\frac{37 \epsilon ^2}{294}$      & $\frac{\epsilon}{15}+\frac{13 \epsilon ^2}{150}$\\
$\beta_3$&$-\frac{2\epsilon}{7}-\frac{229 \epsilon ^2}{735}$    & $-\frac{58 \epsilon ^2}{525}$ & $-\frac{37\epsilon}{105}-\frac{227 \epsilon ^2}{1470}$        & $\frac{\epsilon}{15}-\frac{7 \epsilon ^2}{150}$\\
$\beta_4$&$\frac{8\epsilon}{21}-\frac{103 \epsilon ^2}{735}     $       & $0$                                                   & $\frac{11\epsilon}{35}-\frac{229 \epsilon ^2}{2450}$  & $\frac{\epsilon}{15}-\frac{7 \epsilon ^2}{150}$\\
\noalign{\vskip 1ex} \hline
\multicolumn{5}{c}{Thin prolate particles ($\lambda\to\infty$)}\\
& Total & Unsteady & Convective & Particle \\
$\beta_1$& $\frac{7}{30\log2\lambda-45}$ & $\frac{1}{8\log2\lambda-12}$& $\frac{13}{120\log2\lambda-180}$&$0$\\
$\beta_2$& $\frac{1}{10\log2\lambda-15}$ & $\frac{1}{8\log2\lambda-12}$& $\frac{1}{20\log2\lambda-60}$&$0$\\
$\beta_3$& $0$ &$0$ & $0$&$0$\\
$\beta_4$& $0$ &$0$ & $0$&$0$\\
\end{tabular}
\end{ruledtabular}
\end{table}

The expressions for the contributions from fluid inertia are very lengthy and not particularly instructive. We therefore present the full result graphically as function of aspect ratio $\lambda$ in \Figref{pof_betas}. In addition we give the asymptotic behavior of all contributions to $\beta_\alpha$ in \Tabref{asymptotic} in three limiting cases: 
thin oblate particles ($\lambda\to0$), 
thin prolate particles ($\lambda\to\infty)$, and nearly spherical
particles. For nearly spherical particles we define a small parameter
$\epsilon$ as follows
\begin{align}    
\lambda &= \frac{1}{1-\epsilon} \quad \mbox{for prolate spheroids ($\epsilon>0$)}\,,\nn\\
\lambda &=1+\epsilon\quad \mbox{for oblate spheroids ($\epsilon < 0$)}\,.\nn
\end{align}    
The asymptotic results 
for $\lambda \to 0$, $\lambda \to \infty$, and for $|\epsilon| \to 0 $
are shown as red dashed lines in \Figref{pof_betas}.
\begin{figure*}
\includegraphics[width=\textwidth]{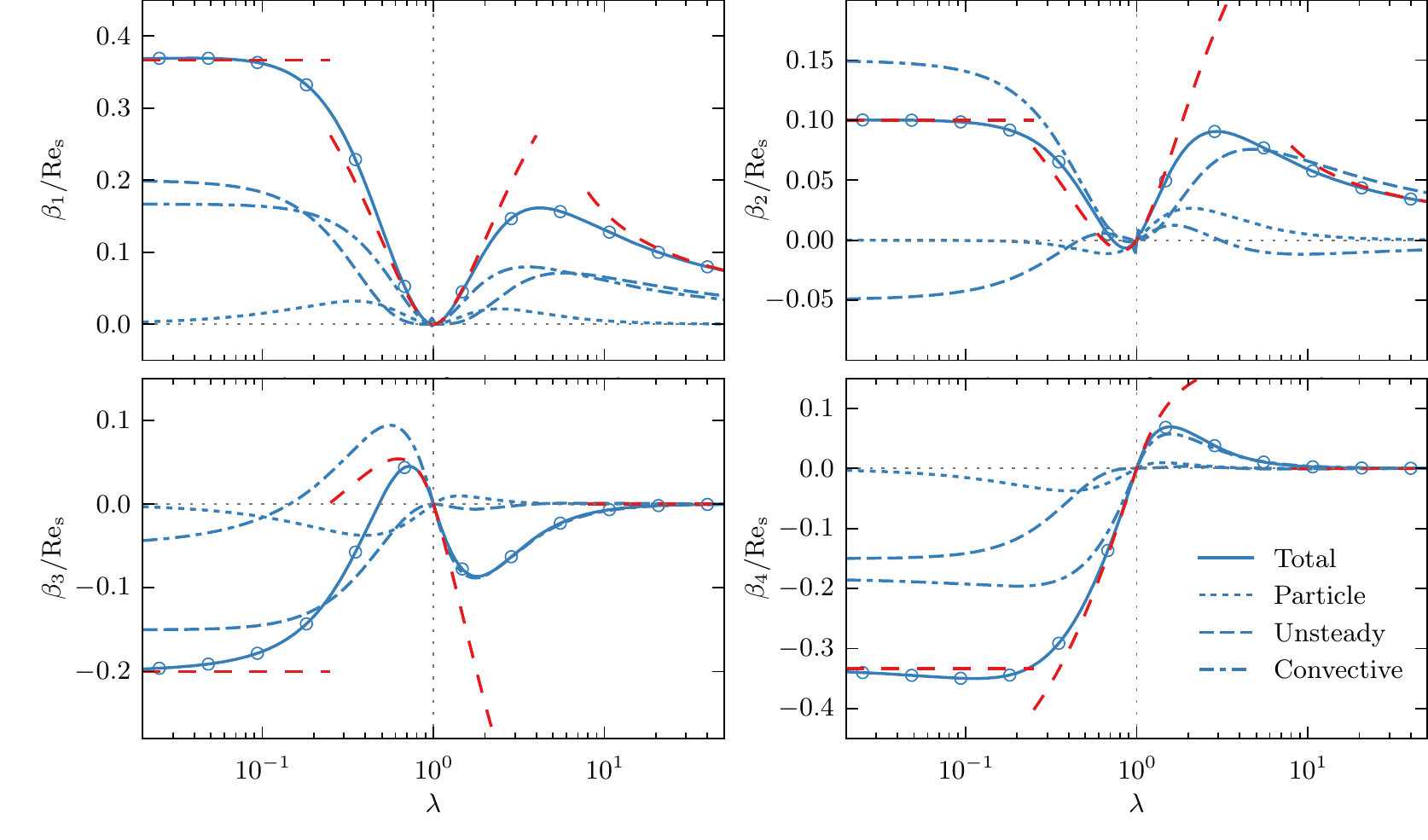}
\caption{\figlab{pof_betas}  {\em Color online.}
Coefficients $\beta_\alpha$ in Eq. (\ref{eqn:effective_spherical}), $\alpha=1..4$ as a function of particle aspect ratio $\ar$ for $\Reys=\St$. 
Solid line shows the sum of all contributions. The other curves show the partial contributions from particle inertia (dotted), unsteady fluid inertia (dashed) and convective inertia (dash-dotted). The red dashed lines show the asymptotic solutions in \Tabref{asymptotic} (first column). Circular markers show result of numerical integration of \Eqnref{omega1} for certain values of $\lambda$.}
\end{figure*}

\subsection{Linear stability analysis at infinitesimal $\Reys$}
\begin{figure}
\includegraphics{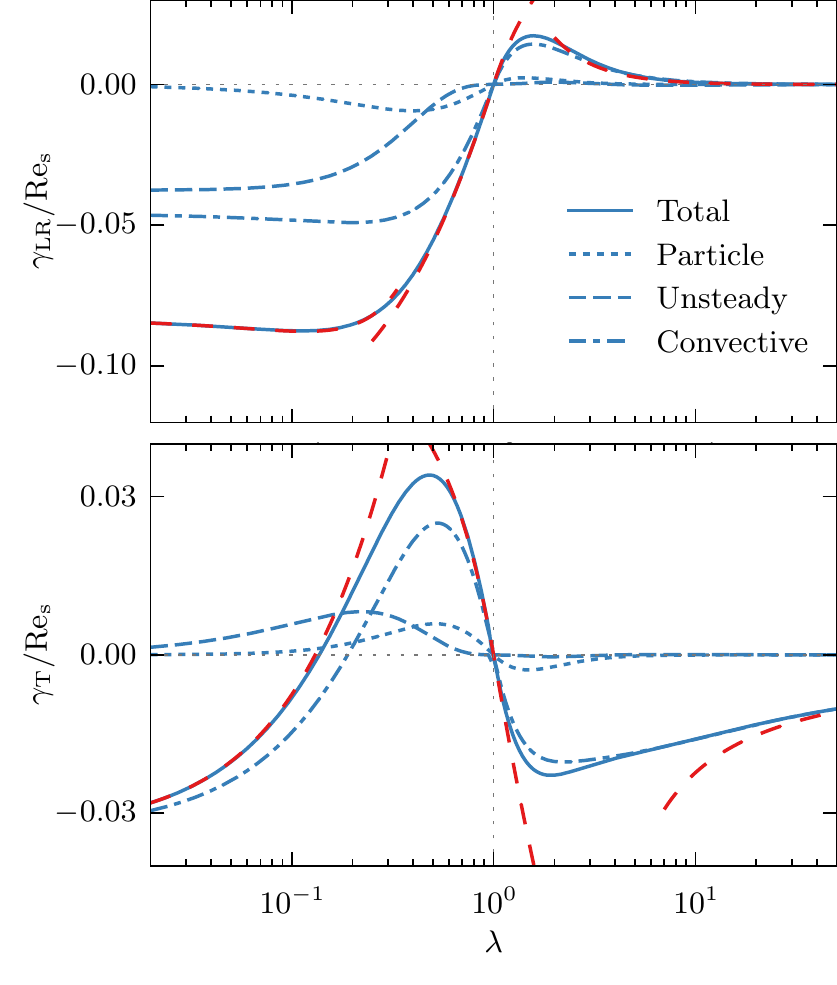}
\caption{\figlab{pof_gammas} {\em Color online.} Stability exponents of log-rolling (top panel) and tumbling (bottom panel) as a function of particle aspect ratio $\ar$ for infinitesimal $\Reys=\St$. Solid line shows the sum of all contributions. The other curves show the partial contributions from particle inertia (dotted), unsteady fluid inertia (dashed) and convective inertia (dash-dotted). Red dashed lines show asymptotic results Eqns.~(\ref{eqn:asymptfirst}-\ref{eqn:asymptlast}).}
\end{figure}
The effective equations of motion \eqnref{effective_spherical} have two special polar angles $\theta$ across which no orbit may pass,
regardless of the values of $\beta_\alpha$. These angles are $\theta = 0$ (the vorticity direction) and at $\theta=\pi/2$ (the flow-shear plane). In the Jeffery dynamics ($\Reys=\St=0$) the two orbits are called \lq log-rolling{\rq} and \lq tumbling{\rq}, and they are both marginally stable, just like all other Jeffery orbits. When the $\beta_\alpha$ are non-zero but infinitesimal, the log-rolling and tumbling Jeffery orbits still exist for any finite aspect ratio, but their stabilities change.

We quantify how particle and fluid inertia 
lift the degeneracy of the Jeffery orbits by computing the stability exponents $\gamma$ for the log-rolling ($\gamma_{\rm LR}$) and tumbling ($\gamma_{\rm T}$) orbits. The stability exponent is the exponential growth rate over one period of the orbit:
\begin{align}
\gamma &= T_p^{-1} \lim_{\delta\theta_0 \to 0}\log\, |\delta\theta(T_p)/\delta\theta_0|
		= T_p^{-1} \int_0^{-2\pi}\frac{\rd \varphi}{\dot\varphi}\frac{\partial \dot\theta}{\partial\theta}\,,
\end{align}
where $T_p = 4\pi/\sqrt{1\!-\!\Lambda^2}$ is the Jeffery period. As $\Reys\!\to\!0$ we find
\begin{align}
\label{eq:gam}
\gamma_{\rm T} &=-\frac{\beta_4}{4}\!+\!\frac{1\!-\!\sqrt{1\!-\!\Lambda^2}}{4\Lambda^2}(\Lambda\beta_2-\beta_1)\,,\quad
\gamma_{\rm LR} = \frac{\beta_4}{4}\,.
\end{align}
For $\Reys=\St$ these two exponents are shown as function of particle aspect ratio in \Figref{pof_gammas}. Also shown are their limiting behaviours in the thin oblate limit ($\lambda\to0$)
\begin{align}
	\frac{\gamma_{\rm T}} {\Reys}&\asympt   
	-\frac{1}{30} \!+\!\left(\frac{7}{30}\!-\!\frac{34}{45 \pi }\!+\!\frac{7 \pi }{80}\right) \lambda \eqnlab{asymptfirst}\\
	&\!+\! \frac{\left(-53248\!+\!19200 \pi \!-\!1728 \pi ^2\!-\!1728 \pi ^3\!+\!567 \pi ^4\right) \lambda ^2}{8640 \pi ^2}
	\nn\\
	\frac{\gamma_{\rm LR}}{\Reys}&\asympt  
	-\frac{1}{12}\! +\!\left(\frac{\pi }{80}\!-\!\frac{16}{45 \pi }\right) \lambda\! +\!\left(\frac{5}{12}\!-\!\frac{256}{135 \pi ^2}\!+\!\frac{3 \pi ^2}{320}\right) \lambda ^2\,,\nn
\end{align}
in the nearly spherical limit ($\epsilon\to0$)
\begin{align}
\label{eqn:nearly_spherical_gamma}
	\frac{\gamma_{\rm T}} {\Reys} \asympt -\frac{2 \epsilon }{21} -\frac{59 \epsilon ^2}{1680} ,\quad
	\frac{\gamma_{\rm LR}}{\Reys}\asympt \frac{2 \epsilon }{21}-\frac{103 \epsilon ^2}{2940},
\end{align}
and in the thin prolate limit ($\lambda\to\infty$)
\begin{align}
	\frac{\gamma_{\rm T}} {\Reys} \asympt \frac{1}{45-30 \log 2 \lambda } ,\quad
	\frac{\gamma_{\rm LR}}{\Reys} \asympt \frac{1}{15\lambda^2}. \eqnlab{asymptlast}
\end{align}
Fig.~\ref{fig:pof_gammas} shows that prolate spheroids of all aspect ratios are unstable at the log-rolling position, and stable at the tumbling orbit. For nearly spherical particles there is a bifurcation: log-rolling and tumbling switch stabilities. For oblate spheroids the log-rolling position is stable for any aspect ratio.
\begin{figure}
\includegraphics{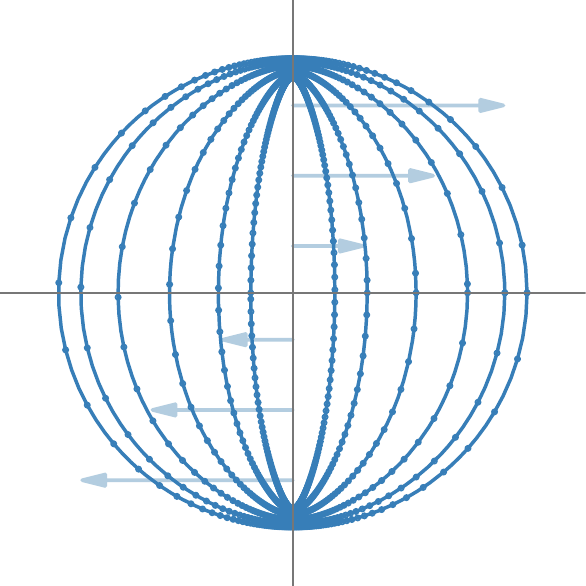}
\caption{\figlab{pof_cycles} {\em Color online.} The shape of the limit cycle for different aspect ratios $\lambda<\lambda_c=1/7.3$. Trajectories are projected onto the unit disk by $[X,Y]=\sqrt{1/(1-n_3)}[n_1,n_2]$ (equal area projection). The tumbling orbit is the unit circle, log-rolling is the center point. The flow-shear directions are indicated in the background. Parameters are, starting from the outermost (tumbling) orbit:
$\lambda=1/7.2$,
$1/7.4$,
$1/8$,
$1/10$,
$1/15$ and
$1/25$.
Data created by numerically integrating \Eqnref{effective_spherical} with $\Reys=10^{-2}$. Markers are spaced equally in time.}
\end{figure}

For oblate particles there is a second bifurcation at 
$\lambda_c \approx 1/7.3$ where the tumbling orbit becomes stable. 
Clearly, this behavior is caused by the convective inertia of the fluid 
(see the dash-dotted line in Fig.~\ref{fig:pof_gammas}).
For sufficiently oblate particles both log-rolling and tumbling orbits are stable, and the long-time dynamics depend on the initial orientation of the particle. Between the two now stable orbits a new unstable limit cycle is born, separating the two basins of attraction. 

\begin{figure}
\includegraphics{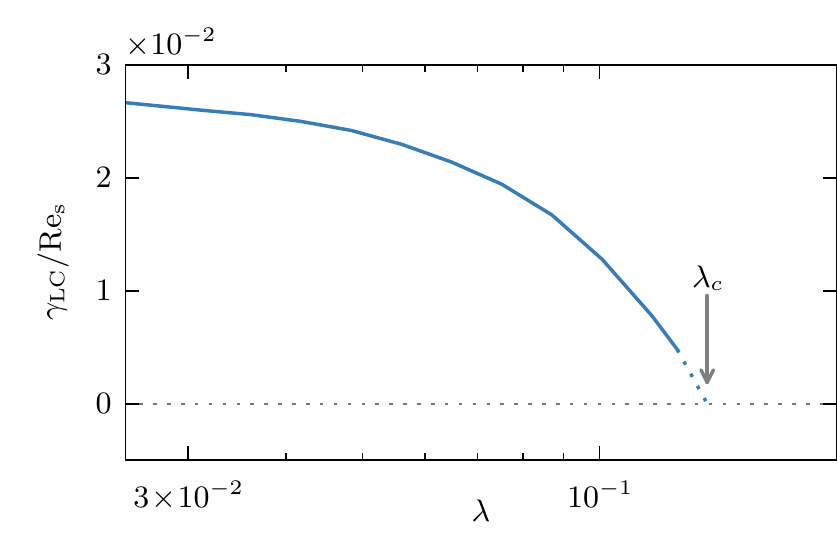}
\caption{\label{fig:pof_gamma_LC} 
{\em Color online.} Stability exponent $\gamma_{\rm LC}$ of the unstable limit cycle as a function of aspect ratio. Computed by numerically
integrating Eqs. (\ref{eqn:effective_spherical}) for $\Reys=0.05$ (solid line). 
The limit cycle bifurcates at $\lambda_{\rm c}$, indicated by an arrow and the dotted continuation of the numerical result.}
\end{figure}

Fig. \ref{fig:pof_cycles} shows how the shape of this limit cycle depends upon
the particle aspect ratio. Close to the bifurcation the limit cycle lies in the neighbourhood of the tumbling orbit. But as $\lambda \to 0$ the limit cycle approaches the
log-rolling orbit. We have computed the stability exponent of the limit cycle
at infinitesimal $\Reys$ by numerically integrating Eqs.~(\ref{eqn:effective_spherical}). The result is shown in  Fig. \ref{fig:pof_gamma_LC}. We see that $\gamma_{\rm LC}>0$, and its magnitude is of the same
order as that of $\gamma_{\rm T}$.

\section{Discussion}
\label{sec:discussion}
{\em Effective  equation of motion.} Eq.~(\ref{eqn:effective_spherical}) 
is an effective equation of motion for the orientational dynamics of a neutrally buoyant spheroid in a simple shear flow. 
How the dynamics depends upon the particle aspect ratio is determined by four coefficients 
$\beta_1,\ldots,\beta_4$. Fig.~\ref{fig:pof_betas} shows the four functions $\beta_\alpha(\lambda)$.
Limiting behaviours of the $\beta_\alpha$ are tabulated
in  Table \ref{tab:asymptotic}. 
We see that the $\beta$-coefficients tend to zero as $\lambda\to \infty$,
 but they approach constants as $\lambda\to 0$. In both limits
the contribution from particle inertia must tend to zero
because the volume of the particle does.
The effects of fluid inertia vanish as $\lambda \to \infty$
because the particle effectively disappears in the slender-body limit, 
the perturbation caused by the particle decreases as $\asympt 1/\log \lambda$ as 
the asymptotic form in Table~\ref{tab:asymptotic} shows.
We remark that the leading-order term in this asymptotic form makes
a substantial correction to the slender-body theory for aspect ratios
of order $30$.

An oblate particle, on the other hand, always presents
no-slip boundaries to the fluid, with an area of the order of $\asympt a^2$ as $\lambda\to0$.
Therefore the contribution of fluid inertia approaches a constant.
We note that the asymptotic forms of the coefficients $\beta_\alpha$
listed in Table \ref{tab:asymptotic} yield accurate values
for $\lambda < 1/30$ and $\lambda > 30$, as Fig. \ref{fig:pof_betas} shows.

We see in Fig. \ref{fig:pof_betas} that the particle-inertia contribution 
to the coefficients $\beta_\alpha$ is always much smaller than the fluid-inertia contributions. In general both unsteady and convective fluid inertia
contribute, and it would be qualitatively wrong to neglect one of these terms.
This is due to the fact that the timescale of the particle motion is the same as the timescale of the flow, and it raises the
question under which circumstances both effects may matter
for the tumbling of small particles in unsteady flows, and
in particular in turbulence.

{\em Linear stability analysis at infinitesimal {\rm $\Reys$}.}
The stability exponents of tumbling and log-rolling orbits 
are shown in \Figref{pof_gammas}. We find that the log-rolling orbit 
is unstable for prolate spheroids of any aspect ratio, tumbling is stable for prolate spheroids, and no other orbit exist at infinitesimal $\Reys$. 
For moderately oblate particles with aspect ratios $\lambda>\lambda_{\rm c}\approx 1/7.3$ the stabilities are reversed: log-rolling is stable, tumbling is unstable, and no other periodic orbits exist for infinitesimal $\Reys$.
At $\lambda=\lambda_c$ there is 
a bifurcation where an unstable periodic orbit is born close to the tumbling orbit, which in turn becomes stable. As $\lambda$ becomes even smaller, the unstable orbit moves closer to the log-rolling orbit (\Figref{pof_cycles}).
We remark that the asymptotic forms (\ref{eqn:asymptfirst}) 
and (\ref{eqn:asymptlast}) of the stability
exponents yield very accurate approximations for the log-rolling exponent,
save for aspect ratios close to unity. For the tumbling exponent the asymptotes
do not work equally well.

Our results are in agreement with results
of recent DNS studies\cite{qi2003,you2007,huang2012,rosen2014,mao2014}
determining the orientational dynamics of a neutrally buoyant spheroid in a simple shear flow. 
These studies are conducted for a number of different aspect ratios
with shear Reynolds numbers ranging from moderate to large.
At the smallest values of $\Reys$ accessible in the DNS
no stable log-rolling is found for prolate spheroids of any aspect ratio.
For oblate particles with aspect ratio $\lambda=1/5$ DNS show stable log-rolling 
and unstable tumbling at the smallest $\Reys$ 
that were simulated\cite{mao2014}, also in agreement with our results. There are no simulations for particles for $\lambda<\lambda_c$ at small $\Reys$.

Saffman\cite{saffman1956} predicted  that 
log-rolling is stable for nearly spherical prolate particles,
at variance with the behaviour described above.
We do not know why the original calculation fails 
to give the correct stability of log-rolling.
Since no details of the calculation are given it is
difficult to figure out the precise origin of this discrepancy.  Subramanian \& Koch\cite{subramanian2006} also computed the stability of the log-rolling orbit for nearly spherical particles and came to the same conclusion as Saffman, different from ours. We have compared the small-$\epsilon$ limit of our calculation to the results of Ref.~\citenum{subramanian2006} and find that the particle-inertia correction to the equation
of motion agrees, Eqs.~(3.15) and (3.16) in Ref.~\citenum{subramanian2006}.
But the fluid-inertia correction does not satisfy the symmetries of the problem. We believe that this explains the discrepancy.

We have independently calculated the stability of log-rolling
for nearly spherical particles by expanding the particle-angular velocity jointly in $\epsilon$ and $\Reys$, using spherical harmonics as a basis set\cite{candelier2015}.
The results of this calculation agree  to order $\epsilon$ with the
results presented above. 
Further we have checked that the particle-inertia correction 
in Eq.~(\ref{eqn:effective_spherical}) is consistent with the results obtained in Ref.~\citenum{einarsson2014}.
We also compared the slender-body limit of our results to the prediction of Subramanian \& Koch for the dynamics of slender fibres\cite{subramanian2005} and found that the fluid-inertia corrections agree (up to a factor of $8\pi$). 

These observations indicate 
that the results presented in this paper are correct, explain the results of DNS
and resolve the puzzle concerning the stability of log-rolling of spheroids in a simple shear at small $\Reys$.

{\em A new benchmark for DNS at small $\Reys$.}
Recently a number of groups have developed DNS codes based on the 
lattice Boltzmann method to simulate the dynamics of particles in 
flows\cite{qi2003,huang2012,rosen2014,mao2014}. 
Much effort is spent on validating the model,
studying for instance the effects changing grid size, time step, size of the simulation
box, and so forth. The benchmark adopted is often the question
whether Jeffery orbits are seen
for a neutrally buoyant spheroid in a simple shear at small Reynolds numbers.
But the limit $\Reys=0$ can never be strictly reached in the simulations.
DNS at small values of
$\Reys$ (specifically: in the linear regime), by contrast, 
allow precise comparisons
with the results obtained in this paper. One could for instance 
compare trajectories, stability exponents, and period times. We thus expect that our results can serve as benchmarks for present and future DNS codes.

\section{Conclusions}
\label{sec:conclusions}
In this paper we have derived an effective equation of motion for the orientational dynamics
of a neutrally buoyant spheroid suspended in a simple shear flow. The equation is valid for arbitrary aspect ratios
and to linear order in $\Reys$, at small but finite shear Reynolds numbers.
The effective equation of motion allows us to determine how the degeneracy of the Jeffery orbits is 
lifted by weak inertial effects. We have determined the bifurcations that occur at infinitesimal $\Reys$ as the particle aspect ratio changes. For prolate spheroids log-rolling is unstable, for oblate spheroids it is stable. Tumbling in the shear plane is stable for prolate particles and unstable for nearly spherical oblate particles. For thin disks with aspect ratios $\lambda<1/7.3$, both log-rolling and tumbling are stable. An unstable limit cycle separates the basins of attraction of the periodic orbits.

Our results imply that tumbling and log-rolling orbits survive a finite perturbation whose magnitude depends on the aspect ratio $\lambda$. 
It would be of interest to 
derive a bifurcation diagram in the $\lambda$-$\Reys$-plane for small $\Reys$. We plan to determine
how the small-$\Reys$ region of this diagram connects to the intricate bifurcation
patterns that were found by Ros\'e{}n, Lundell \& Aidun\cite{rosen2014} at larger shear Reynolds numbers.
We expect that the results summarised here can guide numerical computations with the lattice Boltzmann method that become difficult at small $\Reys$ and large aspect ratios.

\newpage

\appendix
\section{Solutions to Stokes' equation}\applab{stokessolutions}

In this Appendix we solve the steady Stokes' equation for an arbitrarily aligned spheroid in a general linear flow $\ve u^{\infty}=\ma A^\infty\ve r$. The calculation is a special case of the calculation by \citet{jeffery1922}. 
However, instead of the ellipsoidal harmonics that Jeffery used, we employ a finite multipole expansion, following \citet{chwang1975}. The purpose of this Appendix is to derive an explicit closed form expression for the Stokes flow field, suitable for evaluation in the reciprocal theorem. For a more general description of the method we refer to the book by \citet{kim1991}.

{\em Formulation of the problem}.
Stokes' equation reads:
\begin{align}
  \partial_j \partial_j u_i &= \partial_i p\,,\qquad
  \partial_i u_i = 0\,,  \eqnlab{stokes}
\end{align}
with  no-slip boundary conditions on the surface $S$ of the particle
\begin{align}
  u_i &=  \lc_{ijk}\omega_j r_k\quad\mbox{for}\quad \ve r \in S\,. 
\end{align}
Here $\omega_j$ is the angular velocity of the particle.
Furthermore it is assumed that the flow remains unperturbed at
infinitely far away from the particle
\begin{align}
  u_i &=  u_i^\infty \quad\mbox{as}\quad  |\ve r| \to \infty\,.
\end{align}
We solve for the disturbance flow $u_i' = u_i - u_i^\infty$ that 
satisfies Stokes' equation (\ref{eqn:stokes})  with  boundary conditions
\begin{align}
  u_i' &= \lc_{ijk}\omega_j r_k - u_i^\infty\quad\mbox{for}\quad\ve r \in S\,, \quad
  u_i' = 0\quad\mbox{as}\quad |\ve r| \to \infty\,.\eqnlab{disturbance_boundary}
\end{align}
We decompose the linear background flow $u_i^\infty$ into its symmetric and antisymmetric parts, defining the vector $\Omega^\infty_i$ and strain $S_{ij}^\infty$ by 
\begin{align}
u_i^\infty&=A_{ij}^\infty r_j = \lc_{ijk} \Omega_j^\infty r_k + S_{ij}^\infty r_j.
\end{align}
Finally, in terms of the \lq slip angular velocity{\rq} $\Omega_i = \Omega^\infty_i-\omega_i$, the problem to be solved reads
\begin{align}
  \partial_j \partial_j u'_i &= \partial_i p'\,,  \quad
  \partial_i u'_i = 0\,,  \nn \\ 
  u'_i &= -\lc_{ijk}\Omega_j r_k - S_{ij}^\infty r_j\quad\mbox{for}\quad \ve r \in S\,, \nn\\
  u'_i &= 0\quad\mbox{as}\quad|\ve r| \to \infty\,.\eqnlab{disturbance_stokes}
\end{align}

{\em Multipoles}.
We solve \Eqnref{disturbance_stokes} by a finite multipole expansion \cite{chwang1975,kim1991}. 
The multipoles are the Green's function for the Stokes' equation, and its derivatives. In this Appendix we use the shorthand notation $\mathcal G_{ij,k}\equiv\partial_k \mathcal G_{ij}$. The multipoles needed 
to solve for the fluid velocity field around particles in a linear flow are
\begin{align}
  \mathcal G_{ij} &= \frac{\delta_{ij}}{r} + \frac{x_ix_j}{r^3}\,, \nn\\
  \mathcal G_{ij,k} &= -\frac{\delta_{ij}x_k}{r^3} + \frac{\delta_{ik}x_j}{r^3} + \frac{\delta_{jk}x_i}{r^3}-\frac{3x_ix_jx_k}{r^5}\,, \nn\\
  \mathcal G_{ij,ll} &= \nabla^2 \mathcal G_{ij} = \frac{2\delta_{ij}}{r^3} - \frac{6x_ix_j}{r^5}\,,  \nn\\
  \mathcal G_{ij,llk}&= \nabla^2 \mathcal G_{ij,k} =  
  					 -\frac{6}{r^5}\left(\delta_{ij}x_k +\delta_{jk}x_i +\delta_{ik}x_j\right)+\frac{30x_ix_jx_k}{r^7}\,. 
\end{align}
The following two higher-order multipoles are required in the reciprocal theorem. We include them for reference:
\begin{align}
  \mathcal G_{ij,km} &= \frac{\delta_{im} \delta_{jk}}{r^3}+\frac{\delta_{ik} \delta_{jm}}{r^3}-\frac{\delta_{ij} \delta_{km}}{r^3}-\frac{3 x_i x_j \delta_{km}}{r^5} \nn \\
  &\qquad-\frac{3 x_j x_k \delta_{im}}{r^5}  -\frac{3 x_j x_m \delta_{ik}}{r^5}+\frac{3 x_k x_m \delta_{ij}}{r^5}  \nn\\
  &\qquad -\frac{3 x_i x_k \delta_{jm}}{r^5}-\frac{3 x_i x_m \delta_{jk}}{r^5}+\frac{15 x_i x_j x_k x_m}{r^7}\,, \nn\\
  \mathcal G_{ij,llkm}&=\frac{30 x_i x_m \delta_{jk}}{r^7}+\frac{30 x_i x_j \delta_{km}}{r^7}+\frac{30 x_i x_k \delta_{jm}}{r^7} \nn\\
  &\qquad +\frac{30 x_j x_k \delta_{im}}{r^7}+\frac{30 x_j x_m \delta_{ik}}{r^7} +\frac{30 x_k x_m \delta_{ij}}{r^7} \nn\\
  &\qquad -\frac{6 \delta_{im} \delta_{jk}}{r^5}-\frac{6 \delta_{ik} \delta_{jm}}{r^5}-\frac{6 \delta_{ij} \delta_{km}}{r^5} \nn\\
  &\qquad -\frac{210 x_i x_j x_k x_m}{r^9}\,.
\end{align}
Note that we use the ``Oseen tensor'' notation. The Green's function for the Stokes' equation is in fact $G_{ij}=\mathcal G_{ij}/8\pi$. It is convenient to split the dipole contribution $\mathcal G_{ij,k}$ into its antisymmetric 
(\lq rotlet\rq{}) and symmetric (\lq stresslet\rq{}) parts. They are
\begin{align}
  \mathcal G^R_{ij,k} &= \frac{1}{2}\left(\mathcal G_{ij,k} - \mathcal G_{ik,j}\right) = \frac{1}{r^3}\left(\delta_{ik}x_j - \delta_{ij}x_k\right), \nn\\
  \mathcal G^S_{ij,k} &= \frac{1}{2}\left(\mathcal G_{ij,k} + \mathcal G_{ik,j}\right) = \frac{\delta_{kj}x_i}{r^3}-\frac{3x_ix_jx_k}{r^5}.
 \end{align}

{\em Spheroidal multipoles}.
Whereas the flow around a spherical particle may be represented by multipoles anchored at a  single point, representing the flow around a spheroidal particle requires a weighted line distribution of multipoles \cite{chwang1975,kim1991}. We therefore define the \lq spheroidal multipoles{\rq} as the following distributions, note especially the different weights for higher-order multipoles:
\begin{align}
  \mathcal Q^R_{ij,k}(\ve r, \ve n) &= \int_{-c}^c \rd \xi (c^2-\xi^2)\mathcal G^R_{ij,k} (\ve r - \xi \ve n), \nn\\
  \mathcal Q^S_{ij,k}(\ve r, \ve n) &= \int_{-c}^c \rd \xi (c^2-\xi^2)\mathcal G^S_{ij,k} (\ve r - \xi \ve n), \nn\\
  \mathcal Q^Q_{ij,ll}(\ve r, \ve n) &= \int_{-c}^c \rd \xi (c^2-\xi^2)^2\mathcal G_{ij,ll} (\ve r - \xi \ve n).\eqnlab{spheroidalrotletsdef}
\end{align}
The constant $c$ is related to the spheroidal geometry. Prolate and oblate coordinates are obtained by rotating an ellipse around its major or minor axis. We call the distance between the foci of the underlying ellipse $d$, and then $c=d/2$ for prolate coordinates, and $c=id/2$ for oblate coordinates (see definition of coordinate systems in \Appref{spheroidalcoords}.) 

In order to write down explicit tensor expressions for the spheroidal multipoles we introduce the integrals $I^n_m$, $J^n_m$ and $K^n_m$ by
\begin{align}
  I^n_m &= \int_{-c}^c \rd \xi \frac{\xi^n}{|\ve r - \xi \ve n|^m}, \nn\\
  J^n_m &= c^2 I^n_m - I^{n+2}_m, \nn\\
  K^n_m &= c^2 J^n_m - J^{n+2}_m = c^4 I^n_m - 2c^2I^{n+2}_m + I^{n+4}_m.
\end{align}
The spatial variation of the functions $I^n_m$ depends upon $|\ve r|^2$ and $\ve r \bcdot \ve n$ only. Further properties and evaluation of the integrals are discussed in \Appref{spheroidalintegrals}. With $J^n_m$ and $K^n_m$ we express the spheroidal multipoles explicitly, for example the spheroidal rotlet:
\begin{align}
  \mathcal Q^R_{ij,k}(\ve r, \ve n) &= 
  \int_{-c}^c \rd \xi  \frac{c^2-\xi^2}{|\ve r - \xi \ve n|^3} 
 \left[\delta_{ik}(r_j - \xi n_j) - \delta_{ij}(r_k-\xi n_k)\right] \nn\\
  &= (\delta_{ik}r_j - \delta_{ij}r_k)J^0_3 +
    (\delta_{ij}n_k - \delta_{ik}n_j)J^1_3\,. \nn
\end{align}
The integrals $I^n_m$ play the same part in spheroidal geometry as does $1/r^m$ in spherical geometry. The spheroidal stresslet and quadrupole are given by
\begin{align}
  \mathcal Q^S_{ij,k}(\ve r, \ve n) 
  &=\delta_{kj}x_i J^0_3 - \delta_{kj}n_i J^1_3 - 3 r_ir_jr_k J^0_5 - \delta_{jk}n_i J^1_3\nn\\
   &\qquad  + 3(n_ir_jr_k+n_jr_ir_k+n_kr_ir_j)J^1_5 \nn\\
   &\qquad -3(r_in_jn_k+r_jn_in_k+r_kn_in_j)J^2_5\nn\\
   &\qquad+3n_in_jn_k J^3_5\,,
\end{align}
\begin{align}
  \mathcal Q^Q_{ij,llk}(\ve r, \ve n) 
  &= -6(\delta_{jk}r_i+\delta_{ik}r_j+\delta_{ij}r_k)K^0_5 + 30 r_i r_j r_k K^0_7 \nn\\
  &\qquad +6(\delta_{jk}n_i+\delta_{ik}n_j+\delta_{ij}n_k)K^1_5 \nn\\
  &\qquad -30(r_ir_jn_k+r_ir_kn_j+r_jr_kn_i)K^1_7  \nn\\
  &\qquad + 30(n_i n_j r_k + n_i n_k r_j + n_jn_kr_i)K^2_7 \nn\\
  &\qquad- 30 n_in_jn_k K^3_7\,.
\end{align}

{\em Solution by a finite multipole expansion}.
The spheroidal multipoles are functions that satisfy Stokes' equation, and 
a suitable linear combination of them also satisfies the no-slip boundary condition on the surface of a spheroid with symmetry axis $\ve n$. The remaining problem is to determine the coefficients for this linear combination.

Following \citet{kim1991} we use the following ansatz for the disturbance flow field:
\begin{align}
  u'_i &= \mathcal Q^R_{ij,k}\varepsilon_{jkl}\left[
    \left(A^R n_ln_m + B^R(\delta_{lm} - n_ln_m)\right)\Omega_m
    + C^R \varepsilon_{lmn} n_m S_{no}n_o\right] \eqnlab{flowansatz}\\
  & \quad + \left(\mathcal Q^S_{ij,k} + \alpha \mathcal Q^Q_{ij,llk} \right) \nn\\
  &\times\left[\left(A^S n^A_{jklm} + B^S n^B_{jklm} + C^S n^C_{jklm}\right)S^\infty_{lm} 
     - C^R \left(\varepsilon_{jlm}n_kn_m + \varepsilon_{klm}n_jn_m\right)\Omega_l
  \right]\,,\nn
\end{align}
where
\begin{align}
  n^A_{jklm} &= (n_jn_k - \frac{1}{3}\delta_{jk})(n_ln_m - \frac{1}{3}\delta_{lm})\,,  \nn\\
  n^B_{jklm} &= n_j\delta_{kl}n_m + n_k\delta_{jl}n_m + n_j\delta_{km}n_l + n_k\delta_{jm}n_l 
  - 4n_jn_kn_ln_m\,,  \nn\\
  n^C_{jklm} &= -\delta_{jk}\delta_{lm} + \delta_{jl}\delta_{km} + \delta_{kl}\delta_{jm}  \nn\\
        &\qquad+ \delta_{jk} n_ln_m + \delta_{lm}  n_jn_k 
          - n_j\delta_{kl}n_m - n_k\delta_{jl}n_m \nn\\
          &\qquad  - n_j\delta_{km}n_l- n_k\delta_{jm}n_l + n_jn_kn_ln_m\,.
\end{align}
Given the ambient strain $S^\infty_{ij}$, and angular slip velocity $\Omega_i = \Omega^f_i-\omega^p_i$ we must determine seven unknown scalars, which may depend upon the particle shape: $A^R$, $ B^R$, $ C^R$, $ A^S$, $ B^S$, $ C^S$, and $\alpha$. When the coefficients are known, \Eqnref{flowansatz} is the sought Stokes solution.

In order to match the linear boundary condition \Eqnref{disturbance_boundary} we need the combinations of $J^n_m$ and $K^n_m$ in the ansatz to be constant on the particle surface, much like the scalar function $1/r^m$ is in spherical geometry.

Upon examination, the functions $J^0_3$ and $K^0_5$ are constant on the spheroidal surface. Further, the functions $J^1_3$ and $K^1_5$ can be written as $J^1_3=n_jr_j J'^1_3$ and $K^1_5=n_jr_jK'^1_5$, where $J'^1_3$ and $K'^1_5$ are constant on the spheroidal surface. The remaining spheroidal functions $J^n_5$ and $K^n_7$ which appear in the ansatz \eqnref{flowansatz} are more complicated. However, it turns out that they appear only in the combinations $J^n_5 - 10 \alpha K^n_7$. We therefore choose
\begin{align}
  \alpha = \frac{J^0_5}{10 K^0_7}\evalat{\text{surface}} = \frac{1}{8 \left(\lambda ^2-1\right)}\,.
\end{align}
With this choice of $\alpha$ it holds that, on the surface of the spheroid, 
\begin{align}
  J^0_5 - 10 \alpha K^0_7 &= 0\,,\nn\\
  J^1_5 - 10 \alpha K^1_7 &= 0\,, \nn\\
  J^2_5 - 10 \alpha K^2_7 &= \frac{1}{3}J'^1_3-2 \alpha K'^1_5\,, \nn\\
  J^3_5 - 10 \alpha K^3_7 &= \frac{5}{3}J^1_3-\frac{42}{3}\alpha K^1_5\,,
\end{align}
both for prolate and oblate spheroids.

In order to extract the six independent equations for the six remaining coefficients we exploit that the boundary condition must be satisfied for any choice of $n_j$, $\Omega_j$ and $S^\infty_{jk}$. First, with $S^\infty_{jk}=0$ we contract \Eqnref{disturbance_boundary} with $n_i$, $\lc_{ijk}n_j\Omega_k$ and $\lc_{ipq}n_q\lc_{pjk}n_j\Omega_k$. Secondly, with $\Omega_j=0$, we contract \Eqnref{disturbance_boundary} with $n_i$, $S^\infty_{ij}n_j$ and finally $\lc_{ijk}n_j S^\infty_{kl}n_l$. These six equations together have only one solution. We tabulate the resulting expressions for both oblate and prolate spheroids in \Tabref{ARBRCR}.

Computing the torque on a body due to this flow is straightforward, because by construction\cite{chwang1975} the torque on a body due to the rotlet flow $u_i=\mathcal G^R_{ij,k}\varepsilon_{jkl}A_l$ is $T^R_l = -16\pi A_l$, where $A_l$ is the rotlet strength.
The minus sign is due to the fact that the 
torque is exerted on body by the flow. To compute the torque from the spheroidal rotlet \eqnref{spheroidalrotletsdef} we linearly superpose the contributions from all the contained rotlets. The torque from the flow $u_i=\mathcal Q^R_{ij,k}\varepsilon_{jkl}B_l$ is therefore
\begin{align}
  T_l &= -16\pi \int_{-c}^c \rd \xi (c^2-\xi^2)B_l
 -\frac{64\pi c^3}{3}B_l \equiv c_\xi B_l\,.
\end{align}
The factor $c_\xi$ 
depends only on the aspect ratio of the particle (see \Tabref{ARBRCR}). 

  \begin{table}[p]
  \caption{\tablab{ARBRCR} Coefficients for Stokes-flow solutions and moments of inertia for prolate and oblate spheroids.
  These coefficients are collected in the book by Kim \& Karrila\cite{kim1991}. We tabulate them here for convenience, and  because our conventions differ slightly from those adopted in Ref.~\citenum{kim1991}. We remark that some of the 
coefficients tabulated here assume imaginary values. All physical quantities come out to be real-valued.}
  \begin{tabular}{lcr}
  \hline
  \hline
  \noalign{\vskip 1ex}
  \multicolumn{3}{c}{Expressions common to both prolate and oblate spheroids}\\[1ex]
  $\alpha=\frac{1}{8 \left(\lambda ^2-1\right)}$&
  $A^R=\frac{\sqrt{\lambda ^2-1}}{4 \left(C-\lambda ^3+\lambda \right)}$&
  $B^R=\frac{\sqrt{\lambda ^2-1} \left(\lambda ^2+1\right)}{4 \left(-2 C \lambda ^2+C+\lambda ^3-\lambda \right)}$\\[4ex]
  $C^R=\frac{\left(\lambda ^2-1\right)^{3/2}}{4 \left(-2 C \lambda ^2+C+\lambda ^3-\lambda \right)}$&
  $A^S=\frac{\left(\lambda ^2-1\right)^{3/2}}{4 \left(2 C \lambda ^2+C-3 \lambda ^3+3 \lambda \right)}$&
  $C^S=\frac{\left(\lambda ^2-1\right)^{3/2}}{2 \left(3 C+2 \lambda ^5-7 \lambda ^3+5 \lambda \right)}$\\[4ex]
   &
  $B^S=-\frac{\left(\lambda ^2-1\right)^{3/2} \left(C \lambda +\lambda ^4-3 \lambda ^2+2\right)}{8 \left(-2 C \lambda ^2+C+\lambda ^3-\lambda \right) \left(-3 C \lambda +\lambda ^4+\lambda ^2-2\right)}$
    & \\[4ex]
  \hline
  \multicolumn{3}{c}{
  \begin{tabular}{lC{5cm}C{5cm}}
\multicolumn{3}{c}{Expressions particular to prolate and oblate spheroids}\\[1ex]
& Oblate ($\lambda<1$) & Prolate ($\lambda>1$) \\[1ex]
$C$ & 
$\displaystyle-\sqrt{1-\lambda ^2} \cot ^{-1}\left(\displaystyle\frac{\lambda }{\sqrt{1-\lambda ^2}}\right)$ &
$\displaystyle\sqrt{\lambda ^2-1} \coth ^{-1}\left(\displaystyle\frac{\lambda }{\sqrt{\lambda ^2-1}}\right)$ \\[4ex]
$d$ & $2 \displaystyle\sqrt{1-\lambda ^2}$ & $\displaystyle\frac{2 \sqrt{\lambda ^2-1}}{\lambda }$\\[4ex]
$c$ & $\displaystyle\frac{id}{2} $&  $\displaystyle\frac{d}{2}$  \\[4ex]
$c_\xi$  & $\displaystyle\frac{64}{3} i \pi  \left(1-\lambda ^2\right)^{3/2}$& $-\displaystyle\frac{64 \pi  \left(\lambda ^2-1\right)^{3/2}}{3 \lambda ^3}$ \\[4ex]
$A^I$ & $\displaystyle\frac{8 \pi  \lambda }{15}$ & $\displaystyle\frac{8 \pi }{15 \lambda ^4}$\\[4ex]
$B^I$ & $\displaystyle\frac{4\pi}{15}   \lambda  \left(\lambda ^2+1\right)$ & $\displaystyle\frac{4 \pi  \left(\lambda ^2+1\right)}{15 \lambda ^4}$ \\[4ex]
\end{tabular}
}\\
\hline
\hline
  \end{tabular}\\
\end{table}

\section{Spheroidal integrals}\applab{spheroidalintegrals}
In order to solve Stokes' equation and evaluating the volume integrals in the reciprocal theorem we need to solve integrals on the form
\begin{align}
  I^n_m(|\ve r|^2, \ve r\bcdot \ve n) &= \int_{-c}^c \rd \xi \frac{\xi^n}{|\ve r - \xi \ve n|^m}\,.\eqnlab{Inmdef}
\end{align}
First, when matching boundary conditions we must evaluate the integrals with $\ve r$ on the surface of the spheroidal particle. Second, when evaluating the reciprocal theorem we need to integrate products of two or three $I^n_m$ 
multiplied with the components of the spatial coordinate $\ve r$ over the entire fluid volume outside the particle. Therefore we express the functions $I^n_m$ in a spheroidal coordinate system with symmetry axis $\hat{\ve x}'$ along $\ve n$. 
This is accomplished by a rotational change of variables $\ve r'= \ma R \ve r$, $\hat{\ve x}'=\ma R \ve n$, where the latter equality defines a rotation $\ma R$. The absolute value (distance) between $\ve r$ and $\xi \ve n$ is preserved by a rotation, and the integral is transformed into
\begin{align}
  I^n_m
  &= \int_{-c}^c \rd \xi \frac{\xi^n}{\left[(x'-\xi )^2+(y')^2+(z')^2\right]^{\frac{m}{2}}}\,.
\end{align}
This form is equivalent to the integrals $B_{m,n}$ in \citet{chwang1975}. Geometrically, \Eqnref{Inmdef} represents a line source along the direction $\ve n$. The rotation $\ma R$ places the line source along the $x'$-axis in an auxiliary coordinate system. The result is a function of $|\ve r|^2$ and $x' = \hat{\ve x}'\bcdot\ve r' = \ve n \bcdot \ve r$.

Explicit expressions for $I^n_m$ may be found by direct integration, 
or by a recursion formula\cite{chwang1975}. Since we require only a finite number of integrals, we simply perform the direct integration once and for all and save the result in a table.

Finally, when evaluating the term corresponding to unsteady fluid inertia in the volume integral of the reciprocal theorem, we need to compute the derivatives of $I^n_m$ with respect to the moving vector $\ve n$. By differentiating \Eqnref{Inmdef} we derive the following formula:
\begin{align}
	\frac{\partial}{\partial n_i} I^n_m = mr_iI^{n+1}_{m+2}-mn_iI^{n+2}_{m+2}\,.
\end{align}

\section{Spheroidal coordinates}\applab{spheroidalcoords}

Both oblate and prolate spheroidal coordinates are extensions of a two-dimensional elliptic coordinate system ($\xi_1,\xi_2$). The $\xi_1$-coordinate represents concentric ellipses, while $\xi_2$ represents the corresponding hyperbolas. Their intersections give unique coordinates in the $x$-$y$-plane. An azimuthal angle of revolution $\phi$ denotes the extension into three dimensions.

{\em Oblate spheroidal coordinates}.
Start with the $x$-$y$-plane, and place an ellipse of focal distance $d$ with its minor axis along the $x$-axis. Now revolve the ellipse by $2\pi$ around the $x$-axis to produce an oblate spheroid. Then $\xi_1$ represents concentric oblate spheroidal surfaces, $\xi_2$ represents the corresponding hyperbolic surfaces, and we call $\phi$ the angle of revolution.
The coordinate equations are
\begin{align}
  x &= \frac{d}{2}\xi_1 \xi_2\,, \nn\\
  y &= \frac{d}{2}\sqrt{\xi_1^2+1}\sqrt{1-\xi_2^2}\cos\phi\,, \nn\\
  z &= \frac{d}{2}\sqrt{\xi_1^2+1}\sqrt{1-\xi_2^2}\sin\phi\,.
\end{align}
The coordinate ranges are $0\leq\xi_1<\infty$, $-1\leq\xi_2\leq 1$ and $0\leq \phi\leq 2\pi$, and the volume element $\rd V=\frac{1}{8} d^3 \left(\xi _1^2+\xi _2^2\right)\rd \xi_1\rd \xi_2 \rd \phi$.

In this paper we treat oblate spheroids with dimensionless major axis length unity, and minor axis length $\lambda$. These lengths determine the focal distance $d$ as
\begin{align}
  d&=2 \sqrt{1-\lambda ^2},
\end{align}
and the particle surface is parameterised by
\begin{align}
  \xi_1^{(p)}&= \frac{\lambda }{\sqrt{1-\lambda ^2}}\,.
\end{align}

{\em Prolate spheroidal coordinates}.
Start with the $x$-$y$-plane, and place an ellipse of focal distance $d$ with its major axis along the $x$-axis. Now revolve the ellipse by $2\pi$ around the $x$-axis to produce a prolate spheroid. Then $\xi_1$ represents concentric prolate spheroidal surfaces, $\xi_2$ represents the corresponding hyperbolic surfaces, and we call $\phi$ the angle of revolution.
The coordinate equations are
\begin{align}
  x &= \frac{d}{2}\xi_1 \xi_2\,, \nn\\
  y &= \frac{d}{2}\sqrt{\xi_1^2-1}\sqrt{1-\xi_2^2}\cos\phi\,, \nn\\
  z &= \frac{d}{2}\sqrt{\xi_1^2-1}\sqrt{1-\xi_2^2}\sin\phi\,,
\end{align}
The coordinate ranges are $1\leq\xi_1<\infty$, $-1\leq\xi_2\leq 1$ and $0\leq \phi\leq 2\pi$, and the volume element $\rd V=\frac{1}{8} d^3 \left(\xi _1^2-\xi _2^2\right)\rd \xi_1\rd \xi_2 \rd \phi$.

In this paper we treat prolate spheroids with dimensionless major axis length unity, and minor axis length $1/\lambda$. These lengths determine the focal distance $d$ as
\begin{align}
  d&= 2\frac{\sqrt{\lambda ^2-1}}{\lambda }\,,
\end{align}
and the particle surface is parameterised by
\begin{align}
  \xi_1^{(p)}&= \frac{\lambda }{\sqrt{\lambda ^2-1}}\,.
\end{align}

\end{document}